# Low-loss high-impedance circuit for quantum transduction between optical and microwave photons

**Yuta Tsuchimoto and Martin Kroner**

Department of Physics, ETH Zurich, Zurich, Switzerland

E-mail: tyuta@phys.ethz.ch

## Abstract

Quantum transducers between microwave and optical photons are essential for long-distance quantum networks based on superconducting qubits. An optically active self-assembled quantum dot molecule (QDM) is an attractive platform for the implementation of a quantum transducer because an exciton in a QDM can be efficiently coupled to both optical and microwave fields at the single-photon level. Recently, the transduction between microwave and optical photons has been demonstrated with a QDM integrated with a superconducting resonator. In this paper, we present a design of a QD-high impedance resonator device with a low microwave loss and an expected large single-microwave photon coupling strength of 100s of MHz. We integrate self-assembled QDs onto a high-impedance superconducting resonator using a transfer printing technique and demonstrate a low-microwave loss rate of 1.8 MHz and gate tunability of the QDs. The microwave loss rate is much lower than the expected QDM-resonator coupling strength as well as the typical transmon-resonator coupling strength. This feature will facilitate efficient quantum transduction between an optical and microwave qubit.

## 1. Introduction

Superconducting qubits working in the microwave domain are among the most promising platforms for quantum information processing at local nodes [1]. Constructing long-distance quantum networks based on such local superconducting processors is of great interest for quantum communications and distributed quantum computing [2]. However, the considerable attenuation and thermal noise that microwave photons experience has prohibited building quantum networks over 10s of meters so far [3]. A straightforward solution for this issue is to exploit optical links for the interconnection, which allow us to enjoy low attenuation and large transmission bandwidth of optical fibres as well as negligible thermal noise in the optical domain.

Converting a microwave photon to an optical photon and vice versa is essential to realize these optical links. Several types of transducers [4-7], such as optomechanical [8-10] and electro-optic systems [11, 12], have been investigated enthusiastically. Each of those transducers has its strength in one of the following features: high efficiency [11, 13]; low-thermal noise [8]; or a moderate transduction bandwidth [14]. For efficient quantum state transfer, satisfying all these features is necessary. However, it is highly challenging to realize all of them on a single device, which has prohibited the demonstration of the quantum state transfer so far. The bottleneck of this roadblock is generally weak single-photon

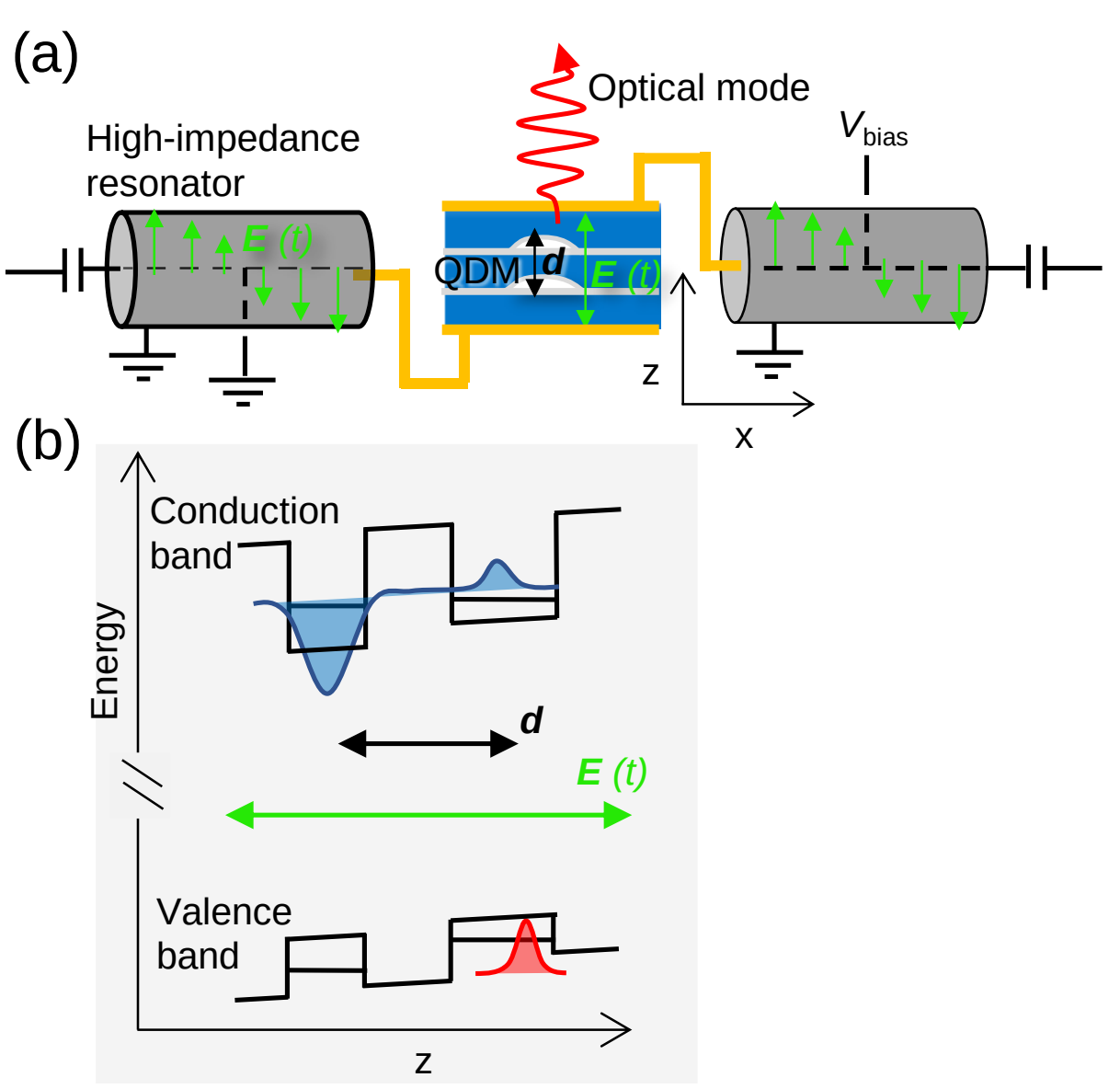

Figure 1. (a) Conceptual schematic of the microwave-optical transducer based on QDM. A MBE grown QDM is transferred onto a pre-patterned low-loss gate structure. The top and back gates are coupled to high-impedance resonators at the position of the electric field antinode, producing a large vacuum electric field ***E*** at the QDM position. (b) The energy band diagram shows an indirect exciton optically excited in the QDM. The blue and red areas represent electron and hole distributions, respectively. The large electric dipole moment ***d*** interacts with the enhanced vacuum field, leading to a large single-photon coupling strength between a microwave photon and the exciton.

coupling between an optical or microwave photon and a medium.

An optically active quantum dot molecule (QDM) [15] is an ideal candidate for the quantum transduction because efficient single-photon coupling to a microwave and optical mode has been demonstrated [16, 17]. We have shown that optical excitation of an exciton creates a large permanent electric dipole moment in a QDM, interacting efficiently with a microwave field in a superconducting coplanar waveguide resonator [16]. The single-photon coupling strength, $g_0$, between an exciton in an InAs/GaAs QDM and a microwave photon has been demonstrated to reach 16 MHz. In addition, thanks to the fast recombination rate $\Gamma_{\rm QD}$ of the exciton, the transduction bandwidth reaches several 100s of MHz.

While for our previous device, the transducer based on a QDM had large single-photon coupling and transduction bandwidth, significant microwave losses were an obstacle for the quantum transduction [16]: the back gate made of a heavily doped GaAs produced a loss of $\kappa/2\pi$ ~ 100 MHz, and the piezoelectric GaAs substrate also led to a loss of several MHz. Moreover, the extensively large $\Gamma_{\rm QD}$, which is beneficial for the large transduction bandwidth, resulted in a relatively small single-photon cooperativity $C_0 = 4g_0/(\Gamma_{\rm QD}+\kappa)\Gamma_{\rm QD} << 1$. This has limited the internal transduction efficiency.

In this paper, we address those issues with an improved device design. In order to reduce the significant losses, we develop a transfer printing technique to incorporate QDMs, hosted in a thin GaAs membrane, into a pre-patterned superconducting resonator on a loss-less substrate. This technique allows us to avoid the large ohmic loss from a heavily doped layer as well as the piezoelectric loss of several MHz from a GaAs substrate.

On the other hand, to achieve a large $C_0$, we employ a high-impedance resonator, which provides a large vacuum fluctuation $\propto \sqrt{Z_{\rm c}} = (L_l/C_l)^{1/4}$ [18-21]. Here $Z_{\rm c}$ is a characteristic impedance of a transmission-line resonator, and $L_l$ and $C_l$ are the inductance and capacitance per length. Since the amount of the vacuum field fluctuation is proportional to the square root of the characteristic impedance, one can enhance the strength of the vacuum field by designing the resonator to exhibit a large $L_l$ and a small $C_l$. The significantly large vacuum field fluctuation enables to enhance $g_0$ as compared to the design based on a typical coplanar waveguide resonator [16]. The expected coupling strength of $g_0/2\pi$ ~ 100s of MHz facilitates achieving $C_0$ ~ 1, improving the internal transduction efficiency to be unity.

Figure 1(a) shows a conceptual schematic of our device. A QDM positioned by transfer printing is sandwiched by low-loss top and back gates made of Au or superconducting Al. The top and back gates are electrically connected to identical high-impedance superconducting resonators at the position of the field antinode, feeding microwave photons from the resonators to the gates. A molecular excitonic state is formed by electron tunnelling between the vertically stacked QDs. This requires that the electronic levels in the two QDs are tuned into or close to resonance. This is achieved by applying a DC bias voltage in the growth/*z* direction (Fig. 1(b)). To this end, two DC bias lines are electrically connected to the top and back gates, respectively. By introducing these bias lines at the field nodes of the two resonators, one can prevent microwave leakage through the bias lines. The optically excited exciton with voltage tunable indirect exciton content [15] has a large electric dipole moment, which efficiently interacts with the microwave electric field oriented in the same direction. For the detailed transduction scheme, see Refs. [16, 22]. Since the design of the circuit is compatible with the integration of typical superconducting qubits, one could incorporate a superconducting qubit into another antinode of the resonator and perform qubit-to-optical photon transduction via the QDM and the resonator bath.

The structure of this paper is as follows. In section 2, we focus on our low-loss high-impedance resonator design without the QD structure. This section includes the basic concepts of our high-impedance resonator design with bias lines, microwave simulation of the circuit, and resonator fabrication. In section 3, we describe how self-assembled QDs are integrated into the high-impedance resonator using a transfer printing technique. In sections 4 and 5, we characterize our fabricated device. In section 4, we show a resonator integrated with QDs and discuss the microwave properties of the devices. In section 5, we investigate the optical properties of gate-tunable QDs transferred on the resonator. Finally, we estimate the coupling strength in section 6.

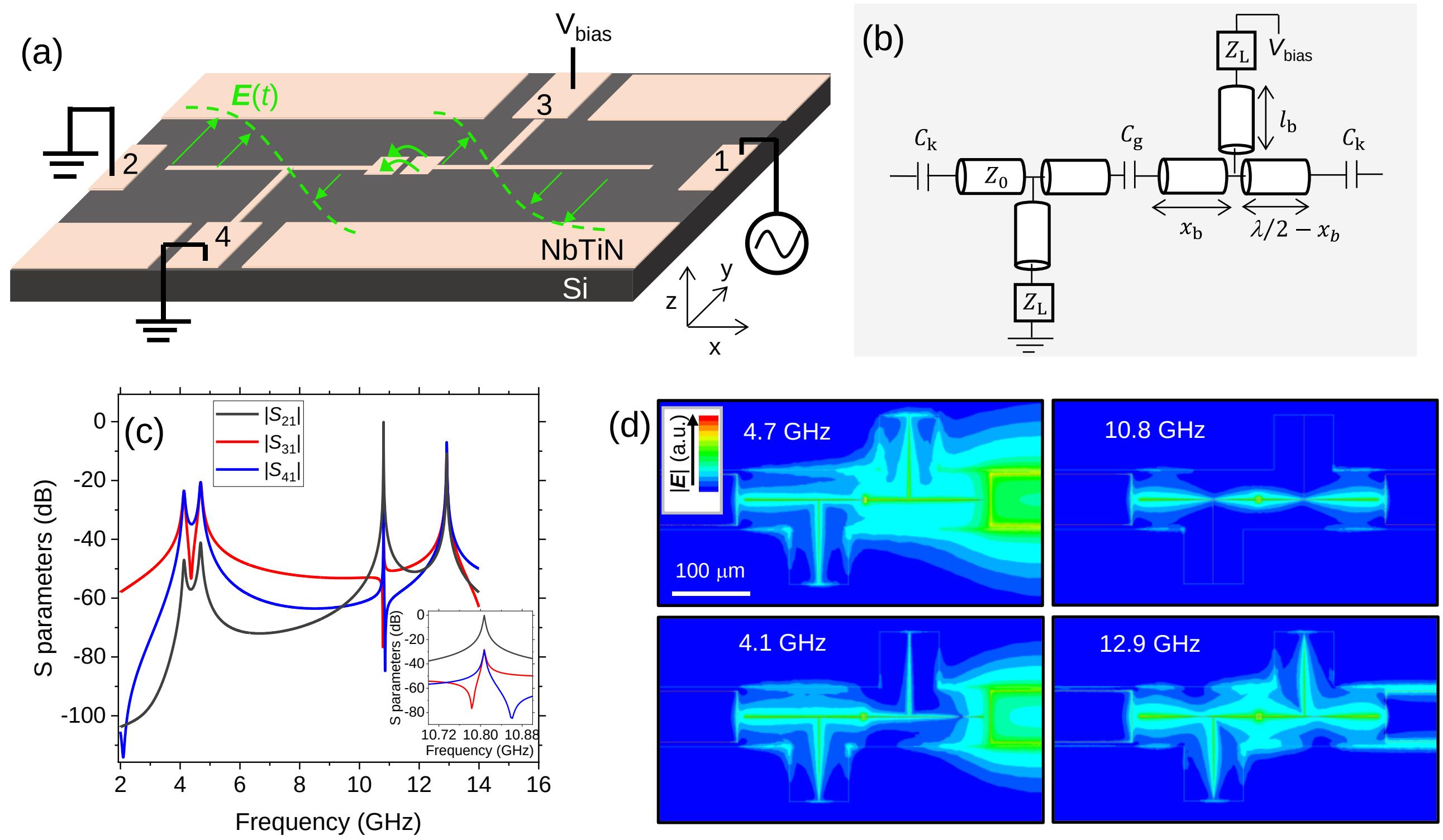

Figure 2. (a, b) Schematic of a coupled nanowire resonator with bias lines and a corresponding circuit diagram (not to scale). The resonator consists of two nanowire resonators with a characteristic impedance $Z_c$. The individual nanowire resonators are coupled with each other by electrodes with a capacitance of $C_g$. The individual resonators are also coupled to feedlines with an external coupling capacitance $C_k$. The dashed green lines show the electric field distribution of the in-phase mode in the resonator. The numbers written on the circuit represent port numbers. (c) Simulated $S$ parameters of the circuit. The subscripts of $|S_{ij}|$ correspond to the port numbers written in (a). The inset shows the enlarged spectrum range around the in-phase mode. (d) shows the electric field distributions of each resonant mode shown in (c).

## 2. A low-loss high-impedance resonator

### *2.1 Desing and simulation of the resonator*

In order to understand how our high-impedance circuit works, we analyze the circuit using electromagnetic field simulations. We designed a high-impedance superconducting resonator based on high-kinetic inductance nanowires made of Niobium Titanium Nitride (NbTiN) [19, 21]. Figure 2(a) shows a schematic of our superconducting circuit consisting of capacitively coupled two half-wavelength nanowire resonators (see Fig. 2(b) for the corresponding circuit diagram). Here, we omitted the QD structure for simplicity: instead of the capacitive coupling by the QD structure, the two identical nanowire resonators are coupled via electrodes of a capacitance $C_g$ at the centre of the circuit. The capacitive coupling between the identical nanowire resonators produces two new coupled modes (in-phase and out-of-phase) because of normal-mode splitting. When the voltages in each nanowire resonator oscillate in phase, the microwave electric field distribution between the nanowires and the ground planes follows the green lines shown in Fig. 2(a). In this case, the capacitively coupled electrodes located at the field antinode generate a strong electric field between them, which can be coupled to an exciton in a QDM via the Stark effect. The nanowire resonators are coupled to feedlines with an external coupling capacitance $C_{k,}$ allowing for transmission measurements to characterize the circuit.

In order to apply a bias voltage to the QDs, we introduced two quarter wavelength bias lines at the electric field nodes of the resonators. Bias lines connected to a resonator must have a filter to prevent microwave leakage, which is generally implemented by a lumped element LC circuit [23] or a half-wavelength resonator loaded by a high impedance lumped element [24]. For high-impedance resonators, implementing typical high-impedance lumped elements is challenging because the filter size should be small to reduce interference originating from the short wavelength of the microwave field. Here we designed the bias lines to be a quarter-wave impedance transformer so that the input impedance of the bias line is given by $Z_{\mathrm{in}} = Z_{\mathrm{c}}^2/Z_{\mathrm{L}}$ [25]. By terminating the high-impedance ($Z_{\mathrm{c}}$) bias line with a low impedance ($Z_{\mathrm{L}}$) port, one can realize a considerably large $Z_{\mathrm{in}}$. Since the bias line is located at the low-impedance point (electric field node) with a local impedance of $Z_{\mathrm{local}}$, microwave leakage can be suppressed by the large impedance mismatch ($Z_{\mathrm{in}} >> Z_{\mathrm{local}}$).

One can also interpret the bias lines as resonators with a fundamental frequency far from that of the coupled nanowire resonator. Since both ends of the bias lines are terminated by low-impedance points (field nodes), the existence of the quarter-wavelength mode is prohibited if the quality factor of the bias lines is high enough, i.e., $Z_{\mathrm{in}} >> Z_{\mathrm{local}}$.

Figure 2(c) shows simulated $S$ parameters of the circuit calculated by the commercial finite-element method field solver ANSYS HFSS. $|S_{ji}|$ represents the transmission from port $i$ to $j$ as labelled in Fig. 2(a), i.e., $|S_{21}|$ shows the transmission of the coupled nanowire resonator, and $|S_{31}|$ and $|S_{41}|$ represent the leakage to the bias ports. We designed the nanowire width and length as $w$ = 70 nm and $l$ = 150 μm, respectively. The distance between the nanowires and the ground planes is 38 μm. Here, the electrode gap is as small as 200 nm to reproduce the capacitance (~ 1fF) of a QD-coupled nanowire device with additional metal gates. We assumed the dielectric constant of the silicon substrate to be $\varepsilon_{\mathrm{Si}} = 11.6$ [26]. In this geometry, we estimated the capacitance of the resonator to be $C_l$ = 40 pF/m using conformal mapping techniques [27]. The sheet kinetic inductance is set to be 83 pH/sq ($L_{kl}$ = 1.2 mH/m), which reproduces a measured $|S_{21}|$ of a coupled nanowire resonator shown in Supplementary S1. From $C_l$ and $L_l \sim L_{kl}$, the characteristic impedance of the nanowires is $Z_c = \sqrt{L_l/C_l} \sim 5500\ \Omega$. The distance between the nanowire and the feedline is 10 um, corresponding to $C_{\mathrm{k}}$ = 0.77 fF calculated by the commercial finite-element solver Ansys Maxwell. The bias lines are terminated by port 3 or 4 with an impedance of $Z_{\mathrm{L}} = 50\ \Omega$.

In the simulated spectrum of $|S_{21}|$, there are two resonances at 10.8 and 12.9 GHz, which correspond to the in-phase and out-of-phase modes, respectively. The in-phase mode produces a strong electric field between the electrodes of each resonator. We located the bias lines at the field nodes of the in-phase mode so that the mode is protected from the leakage. The inset of Fig. 2(c) shows $S$ parameters around the in-phase mode at 10.8 GHz. The weak transmission of $|S_{31}|$ and $|S_{41}|$ indicate the suppressed microwave coupling between the resonators and bias lines. The dispersive line shapes appearing in $|S_{31}|$ and $|S_{41}|$ stem from the interference between the sharp resonance of the transmission mode and the tail of the resonances associated with the bias lines [24]. Figure 2(d) shows the electric field distributions at each resonance. The in-phase mode at 10.8 GHz exhibits the field distribution protected from the leakage due to the decoupling of the bias lines from the resonators.

The other resonances at 4.1 and 4.7 GHz are resonant with the bias lines, leading to significant leakage, as seen in $|S_{31}|$ and $|S_{41}|$ in Fig. 2(c), and the corresponding electric field distributions in Fig. 2(d). In these cases, the loaded quality factors result in $Q_{\mathrm{L}} < 100$, and the vacuum field fluctuations should be weaker because of the larger mode volumes due to the leakage.

### 2.2 Imperfections of the bias lines

If the parameters of the bias lines deviate from the ideal design due to fabrication imperfections, non-negligible losses will appear in the in-phase mode. To examine the tolerance of the resonator loss against the imperfections in the bias lines, we simulated the decay rates of the resonator depicted in Fig. 2(b) while changing the position $x_{\mathrm{b}}$, length $l_{\mathrm{b}}$ of the bias lines, and $C_{\mathrm{g}}$ (see Fig. 2(b) for the definitions of each variable). In order to assess the decay introduced by the bias lines, we simulated a device with and without the bias lines and extracted the decay rate differences $\delta\kappa$. Here, we set a negligibly small external coupling $C_{\mathrm{k}}$ = 10 aF corresponding to an external decay rate of 1 kHz for this simulation. The characteristic and load impedances are again $Z_{\mathrm{c}} = 5500\ \Omega$ and $Z_{\mathrm{L}} = 50\ \Omega$, respectively. For the simulation, we used a commercial package, Ansys electronics desktop circuit simulator.

Figure 3(a) displays $\delta\kappa$ as a function of $x_{\mathrm{b}}$ with $l_{\mathrm{b}} = \lambda/4$ and $C_{\mathrm{g}}$ = 1 fF, where $\lambda$ is the resonance wavelength of a bare single nanowire resonator. For the design assumed here, the electrical field node is located at $x_{\mathrm{b}} \sim 0.185\lambda$ which is slightly shifted to the centre of the circuit compared with the case of a bare single nanowire resonator ($x_{\mathrm{b}} = 0.25\lambda$) because of the presence of $C_{\mathrm{g}}$. When the positions of the bias lines deviate from the nodes (i.e., $x_{\mathrm{b}} \neq 0.185\lambda$), $\delta\kappa$ becomes larger because a larger $Z_{\mathrm{local}}$ leads to a smaller impedance mismatch between $Z_{\mathrm{in}}$ and $Z_{\mathrm{local}}$. This behaviour can be qualitatively described by the position-dependent transmission coefficient:

$$T(x_{\mathrm{b}}) = 1 - \left(Z_{\mathrm{in}} - Z_{\mathrm{local}}(x_{\mathrm{b}})\right)/\left(Z_{\mathrm{in}} + Z_{\mathrm{local}}(x_{\mathrm{b}})\right), \qquad (1)$$

where $Z_{\mathrm{local}}(x_{\mathrm{b}}) = V_0 cos(2\pi/\lambda x_{\mathrm{b}})/I_0 sin(2\pi/\lambda x_{\mathrm{b}})$. Here we defined the voltage and current amplitudes in the resonator as $V_0$ and $I_0$, respectively. To keep $\delta\kappa$ < 1 MHz, the mismatch between the position of the nodes and the bias lines should be within ± 0.025λ corresponding to ± 7.5 μm in our design.

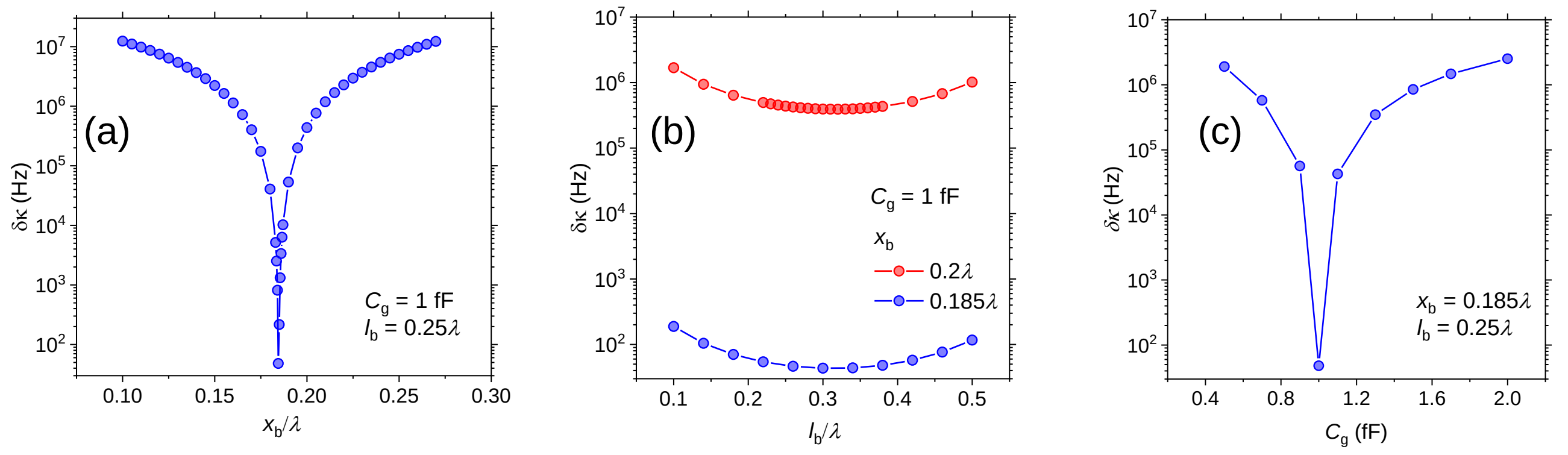

Figure 3. Simulated decay rate changes $\delta\kappa$ of the coupled nanowire resonator due to the leakage through the bias lines depending on some parameters. Throughout the simulations, we set the external coupling negligibly small, $C_k$ = 10 aF. (a) The bias line position ($x_b$) dependence of $\delta\kappa$ at the individual resonator coupling $C_g$ = 1 fF and the bias line length $l_b = 0.25\lambda$. (b) The $l_b$ dependence of $\delta\kappa$ at $C_g$ = 1 fF and $x_b = 0.2\lambda$ (blue) or $0.185\lambda$ (red). (c) The $C_g$ dependence of $\delta\kappa$ at $x_b = 0.185\lambda$ and $l_b = 0.25\lambda$.

Next, we investigated the tolerance against the change of $l_b$. Figure 3(b) shows $\delta\kappa$ as a function of $l_b$ with $C_g$ = 1 fF and $x_b$ = 0.185λ or 0.2λ, corresponding to the positions of the field node or slightly off from the node, respectively. For both cases, $l_b \sim 0.3\lambda$ achieves a minimum of $\delta\kappa$. Here $l_b \sim 0.3\lambda$ is the quarter wavelength of the microwave field on the bias line, which is slightly longer than that of a bare single nanowire resonator (0.25λ) due to the presence of $C_g$. Even if $l_b$ is far away from the quarter wavelength, the increase in $\delta\kappa$ is relatively small compared to the case in Fig. 3(a).

Deviation of an experimentally obtained $C_g$ from a simulated one can also be a significant source of loss. The change in $C_g$ shifts the field node positions, leading to the mismatch between the positions of the field nodes and the bias lines. Figure 3(c) exhibits $\delta\kappa$ as a function of $C_g$ at $x_b = 0.185\lambda$ and $l_b = \lambda/4$, where $C_g$ = 1 fF forms the field nodes at the bias line positions. To keep the loss $\delta\kappa < 1$ MHz, the deviation from $C_g$ = 1 fF should be less than ± 0.5 fF. For a QD-coupled nanowire device shown later, $C_g$ is dominated by a parallel plate capacitor with a thickness of $h_{GaAs} \sim 130$ nm, formed by a top and back gate and GaAs as a medium. In this case, the allowed deviation of the gate length is ± ~ 0.2 μm for a gate size of 1 μm × 1 μm corresponding to $C_g \sim 1$ fF. Typical electron beam lithography (EBL) allows to fabricate the gates with such precision, and careful electrostatic simulation of the device should be helpful to reduce the deviation of the actual $C_g$ from the design.

We remark that the estimated tolerances of $x_b$, $l_b$, and $C_g$ are the results for $Z_L = 50\ \Omega$. To increase the tolerances further, one can reduce $Z_L$, for instance, by introducing additional stubs at the end of the bias lines or simply making a low-impedance transmission line as a bias port with a narrower centre-ground gap and a non-kinetic inductance material.

### *2.3 Fabrication of the coupled nanowire resonator*

We fabricated a coupled nanowire resonator without QDs on a high-resistivity silicon substrate (~ 20 kΩcm). For the superconducting resonators, we sputtered a thin NbTiN film (15 nm) on the silicon substrate. The sheet resistance of the film is 378 Ω/sq at room temperature. We patterned nanowires with a width of ~ 70 nm using EBL followed by reactive ion etching (RIE) in $SF_6/O_2$ plasma. The design parameters of the nanowire resonators and bias lines are mentioned in Section 2.1. We measured $|S_{21}|$ at 55 mK and observed four modes as expected from the simulation. The loaded quality factor of the in-phase mode is $Q_L$ = 7500, corresponding to a decay rate of 1.4 MHz, limited by the external coupling strength. See Supplementary S1 for more details.

## 3. Transfer printing of a QD structure

### *3.1 Fabrication of a QD structure to be transferred*

In this section, we explain how we integrate a self-assembled InAs/GaAs QD structure into the superconducting circuit using a transfer printing technique. First, we grew a GaAs heterostructure hosting vertically stacked InAs/GaAs QDs. The heterostructure was grown on a semi-insulating GaAs substrate using molecular beam epitaxy in Stranski–Krastanov mode. In order to prevent a current through the heterostructure, we introduced current blocking layers made of 18 nm-thick AlAs/GaAs superlattices. The layers are located both above and below the QDs (see Supplementary S2 for the detailed structure). The current blocking layers also block electron tunnelling from QDs to gates that we will fabricate later, expanding the charge-stable region to a broader bias voltage range [28, 29]. This feature allows for tuning the QD exciton resonance frequency across a broader range, facilitating a larger electric dipole moment than the typical QD device with electron tunnelling. The wafer has a sacrificial layer consisting of a ~ 500 nm-thick Al-rich superlattice (AlAs/GaAs) underneath the heterostructure, allowing to lift the heterostructure off from the GaAs substrate [30, 31]. The thickness of the lifted structure with QDs (QD membrane) is $h_{\mathrm{GaAs}}$ = 130 nm for the sample used in this experiment.

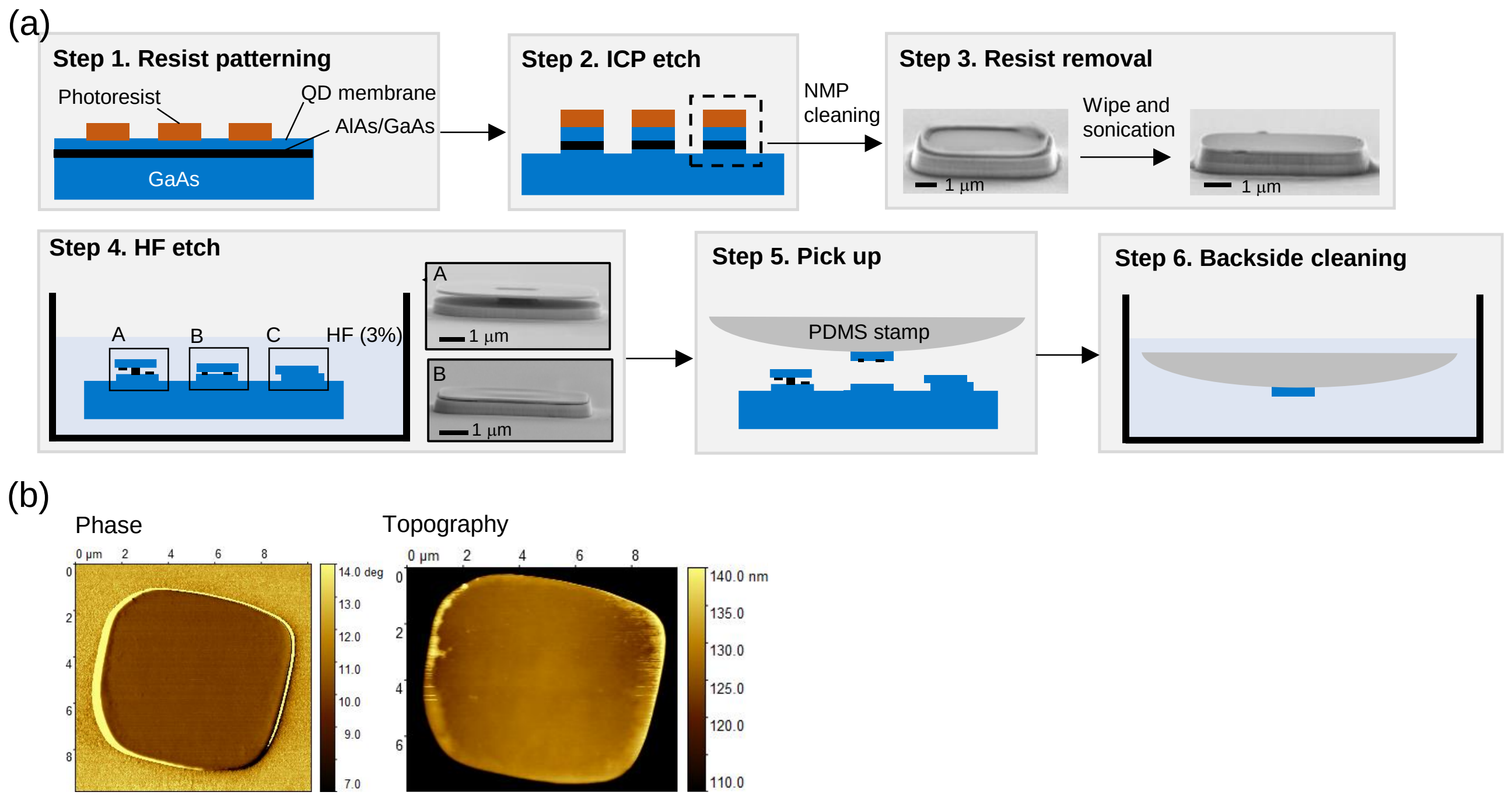

Figure 4. (a) Schematic of the pickup and cleaning processes for transfer printing of QD membranes (not to scale). The processes consist of six steps: Step1. Resist patterning; Step2. ICP etching of GaAs; Step3. Resist removal; Step4. HF etching of the sacrificial layer; Step5. Pickup of a QD membrane; Step6. Cleaning of the back surface. The SEM images show the structures under each processing. (b) Atomic force microscope images (phase and topography) of the back surface of a picked QD membrane on a PDMS stamp. The images were taken by the tapping mode.

Next, we fabricate the QD membranes to be transferred on the coupled nanowire resonator. First, we patterned numerous mesa structures with an area of 6 µm × 6 µm by masking the wafer surface with a photoresist and etching the wafer vertically to the bottom GaAs substrate (see Fig. 4(a) Step 1 and 2). After the etching, we removed the resist by hot N-Methyl-2-Pyrrolidone (NMP) solvent. The left scanning electron microscope (SEM) image in Step 3 shows a typical patterned mesa structure. The dark contrast region in the mesa is the AlAs/GaAs sacrificial layer, and the bright contrast region above the sacrificial layer hosts QDs and current blocking layers. The Al-rich sacrificial layer is made of an AlAs/GaAs superlattice resistant to oxidation. If the sacrificial layer was made of pure AlAs, oxidized AlAs expands drastically after etching and exposing to air, breaking the thin QD membrane. On the top surface of the heterostructure, hardened resist remains along the edge of the structure, which we could not remove by the NMP solvent. We removed it by gently wiping the surface with a cleanroom tissue in deionized water, followed by ultrasonic cleaning (see the right SEM image in Step 3). Here, we avoided plasma cleaning to remove the photoresist masks because the presence of plasma-induced defects nearby the QDs results in fluctuation of the QD energies [32]. The sacrificial layer underneath the QD membrane was removed by soaking it in 3 % hydrofluoric acid (HF) for 2 min. The schematic in Step 4 shows the typical structures after the HF etching. The structure labelled as A displays a QD membrane still partially supported by the sacrificial layer that was not fully etched out. The one labelled as B is a QD membrane where the sacrificial layer has been completely etched out, but the membrane is supported by AlAs/GaAs residues at the back surface. The last structure labelled as C shows a membrane that has completely collapsed onto the GaAs substrate. The SEM images in Fig. 4(a) Step4 are examples for cases A and B. We picked up QD membranes resembling B using an elastomer stamp made of polydimethylsiloxane (PDMS) (Step 5). In the pickup process, we distinguished the three cases by observing thin-film interference patterns with an optical microscope. While the membranes resembling B can easily be picked up without damaging them, the membranes similar to A typically break into pieces when picked up. The ones similar to C cannot be picked up at all because they strongly stick to the substrate. Since we fabricated thousands of mesa structures at once and need only some of them, the relatively low yield of the membranes suitable for pickup (< 10 %) is enough for our experiment. For higher yield, one can fabricate anchors by a photoresist to tether the membranes above the substrate [33]. After pickup, we cleaned the back surface of the membranes by digital etching [34], comprising surface oxidation by 30% $H_2O_2$ followed by acid cleaning with 16 % HCl for 1 min, and additional treatment with 3 % HF for 1 min (Step 6). The digital cleaning with $H_2O_2$ and HCl helps remove contamination on the back surface compared with cleaning with only 3 % HF (See Supplementary S4).

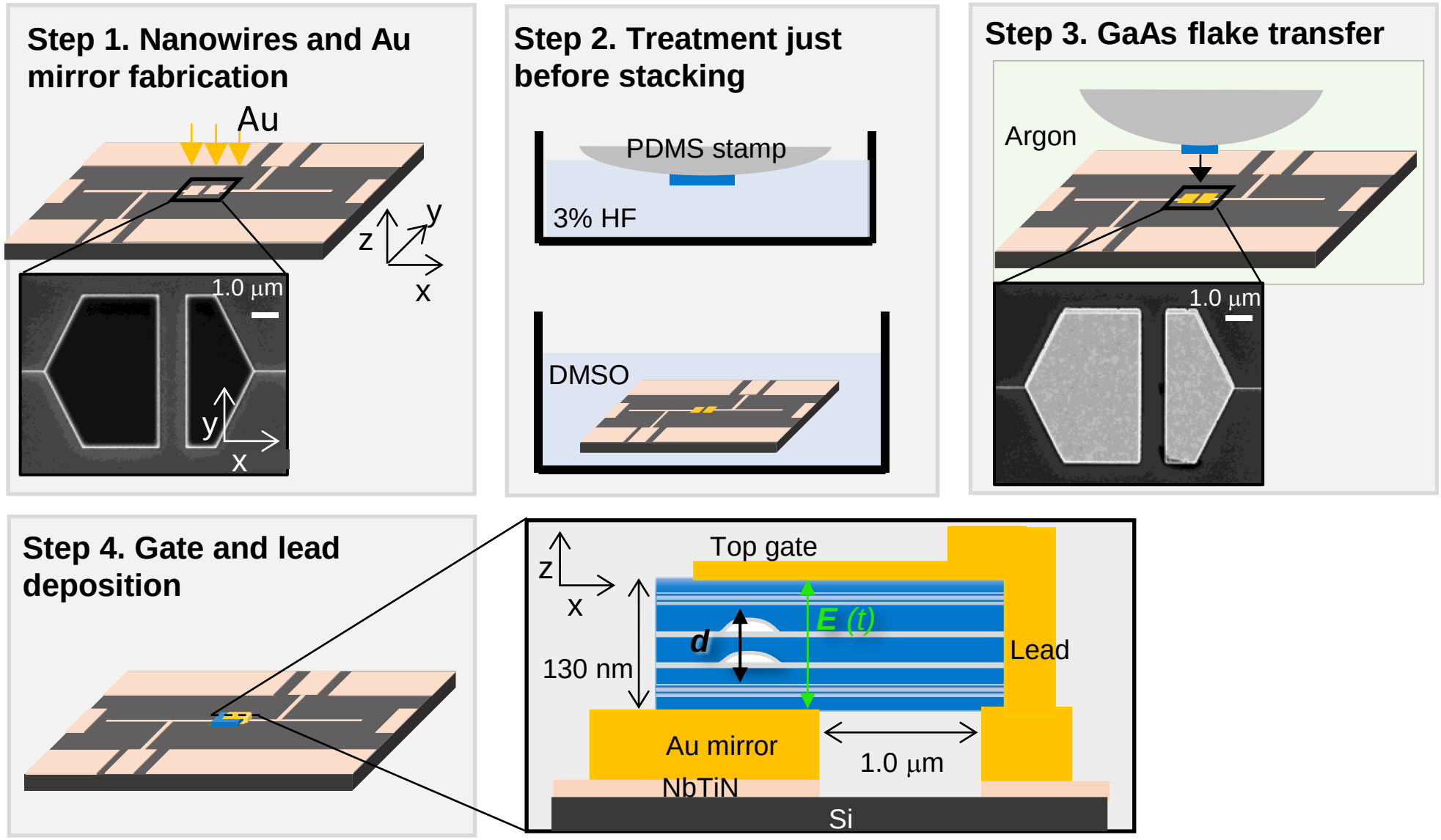

Figure 5. Schematic (not to scale) of the transfer printing processes consisting of Au mirror fabrication (Step 1), cleaning of the circuit and HF treatment for a QD membrane (Step 2), stacking (Step 3), and Ti/Au gate and lead deposition (Step 4). The SEM images show enlarged images of the electrodes connected to the right and left nanowire resonators. The cross-sectional schematic shows a QD membrane sandwiched by an Au mirror and a Ti/Au gate forming a strong vertical electric field at the QDM position.

Figure 4(b) shows atomic force microscope (AFM) images of the cleaned back surface of a QD membrane on a PDMS stamp. The phase images show a clean back surface where no obvious contamination is visible. The slightly visible horizontal stripe patterns in the phase image are artefacts of the scanning AFM tip. We removed these artificial patterns when constructing the topography image by taking the median differences between vertical neighbour pixels. A few small islands with a height of ~ 2 nm are visible in the topography image. There are also bumps with a height of ~ 5 – 10 nm along the edge of the membrane. These structures could be residues such as Aluminum fluoride produced during the HF etching process [31] in Step 4. The root-mean-square (RMS) roughness away from the edge is ~ 0.5 nm, relatively smaller than the previously reported value [31], possibly because of the digital cleaning process (see Supplementary S4).

### *3.2 Transfer process of a QD membrane*

Next, we transfer the QD membrane onto a coupled nanowire circuit. We prepared a bare coupled nanowire resonator as described in section 2.3. On top of the NbTiN electrodes (see the SEM image in Fig. 5 Step 1), we deposited 100 nm-thick Au by electron beam deposition immediately after removing the oxide layer on NbTiN by Ar ion milling (Step 1). The thick Au layer works as an optical mirror and increases the collection efficiency of optical photons from QDs that will be transferred on the Au mirror. The RMS roughness of the Au surface is ~ 1 nm measured by an AFM. Before the stacking process, we cleaned the circuit with a hot Dimethyl sulfoxide (DMSO) solvent and performed treatment for the QD membrane with the 3 % HF solvent (Step 2). After that, we immediately moved them into a glove box (Argon atmosphere). The QD membrane was transferred onto the Au mirror by contacting the membrane to the Au mirror under a microscope and slowly lifting the stamp up in the glove box at room temperature (Step 3). We performed the stacking process more than 10 times and confirmed that the membranes reproducibly adhered well to the Au surface without failure. For the transfer with a high success rate, the surfaces of the Au mirror and the QD membrane should be clean to improve the adhesion. In addition, the size of the contact area and the presence of the HF treatment in Step 2 just before the stacking process seem to be critical (see Supplementary S5 for more details). We emphasize that our stacking process does not require additional adhesion layers, such as self-assembled monolayers [35] or thin polymers [36], between the GaAs membrane and the Au layer.

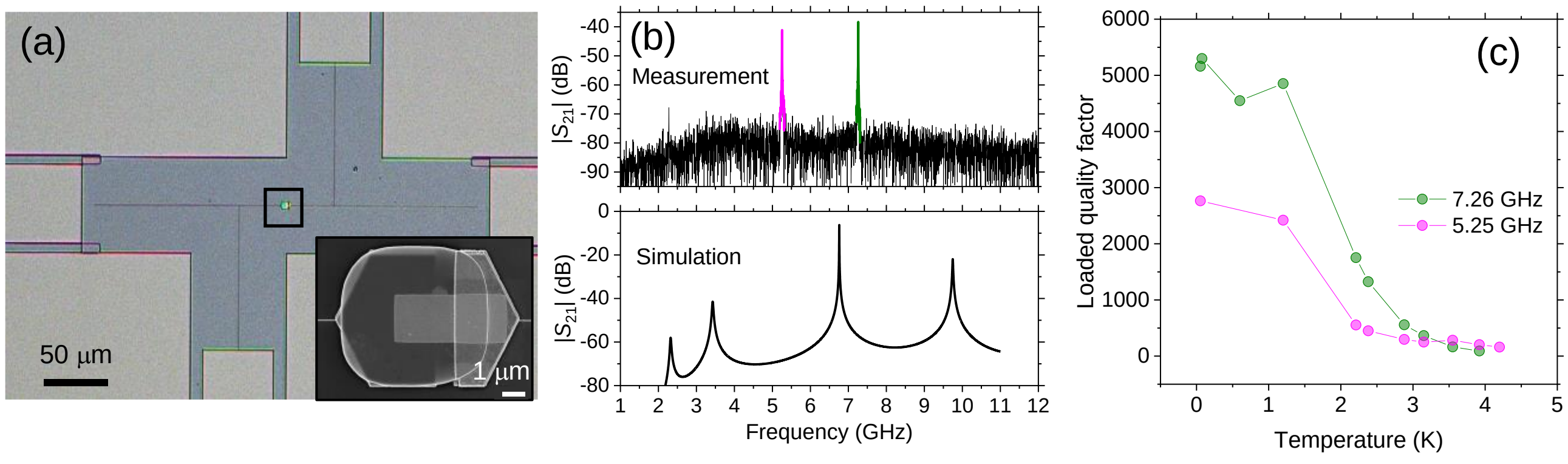

Figure 6. (a) Optical microscope image of a fabricated QD-coupled nanowire resonator and bias lines. The inset displays an SEM image at the black rectangle position, showing a QD membrane with a Ti/Au top gate. (b) Measured (top) and simulated (bottom) $|S_{21}|$ of a QD-coupled nanowire resonator. Two distinctive resonances in the measurement are highlighted by magenta and green. (c) shows the temperature dependence of the loaded quality factor of the modes shown in (b).

After the stacking process, we deposited a 15 nm-thick Ti/Au gate as the top gate on top of the QD membrane. In order to feed microwave photons from the right nanowire resonator to the top gate, we deposited a 100 nm-thick Ti/Au lead on the sidewall and the part of the top gate on the membrane (Step 4). The overlap of the top gate and the bottom Au electrode (mirror) on the left NbTiN electrode generates a strong vertical electric field oriented in the direction of the excitonic permanent dipole moment in a QDM. The lateral capacitive coupling between the electrodes is negligible in this geometry. The thin top gate and the bottom Au mirror on the left NbTiN electrode form an optical cavity, improving the optical outcoupling efficiency.

## 4. Microwave measurement of the QD-nanowire resonator device

Figures 6(a) shows the optical microscope image of a typical QD-coupled nanowire resonator and bias lines with a nanowire width of ~ 40 nm. The inset shows an SEM image of the transferred QD membrane with the Ti/Au gate. We made a relatively large gate overlap (~ 2 μm × 2 μm) between the top gate and the bottom Au electrode for the current test device in order to find the gate-tunable QDs easily in the later optical characterization. To suppress the microwave leakage, we positioned the bias lines to the shifted field node positions because of the relatively large $C_g$ ~ 3 fF. We remark that, for an ideal device, the top gate area should be as small as possible to keep the vacuum fluctuation large. This is because a larger gate area increases the capacitance of the circuit, leading to a lower vacuum fluctuation. One can fabricate a device with the smallest possible gate area by first determining a position of a target QDM in a bulk wafer [37], followed by transfer printing. By patterning the top gate only above the target QDM, the top gate area is minimized. This development remains for future work.

We first performed microwave transmission measurements for a QD-coupled nanowire resonator co-processed with the device shown in Fig. 6(a). Figure 6(b) shows measured $|S_{21}|$ spectrum (top panel) at 1.5 K and simulated one (bottom panel). In the measured $|S_{21}|$, there are two distinctive resonances at 5.25 and 7.26 GHz. We confirmed that those resonances originate from the nanowires by measuring the temperature-dependent resonance frequency shift due to their kinetic inductance change. Comparing the measured resonances with the simulated ones, however, the resonance frequencies do not match very well. This discrepancy might be due to some unexpected imperfections in the nanowire fabrication. For instance, if the width of the superconducting nanowire is locally narrow, such a defect could induce a large inductance leading to a shift of the resonance frequencies. Moreover, the exact frequencies of the in-phase and out-of-phase modes do not only depend on the individual nanowire resonances but also their coupling strength. We deduce that those two distinctive resonances are shifted in-phase (5.25 GHz) and out-of-phase (7.26 GHz) modes for this sample. We emphasize that the discrepancy between the measured and simulated resonance frequencies is particular for this sample and another similar sample exhibits a spectrum well corresponding to the simulation (See Supplementary S6). We believe that the agreement between the actual resonance frequencies and simulated ones can be improved by optimizing the nanowire fabrication process.

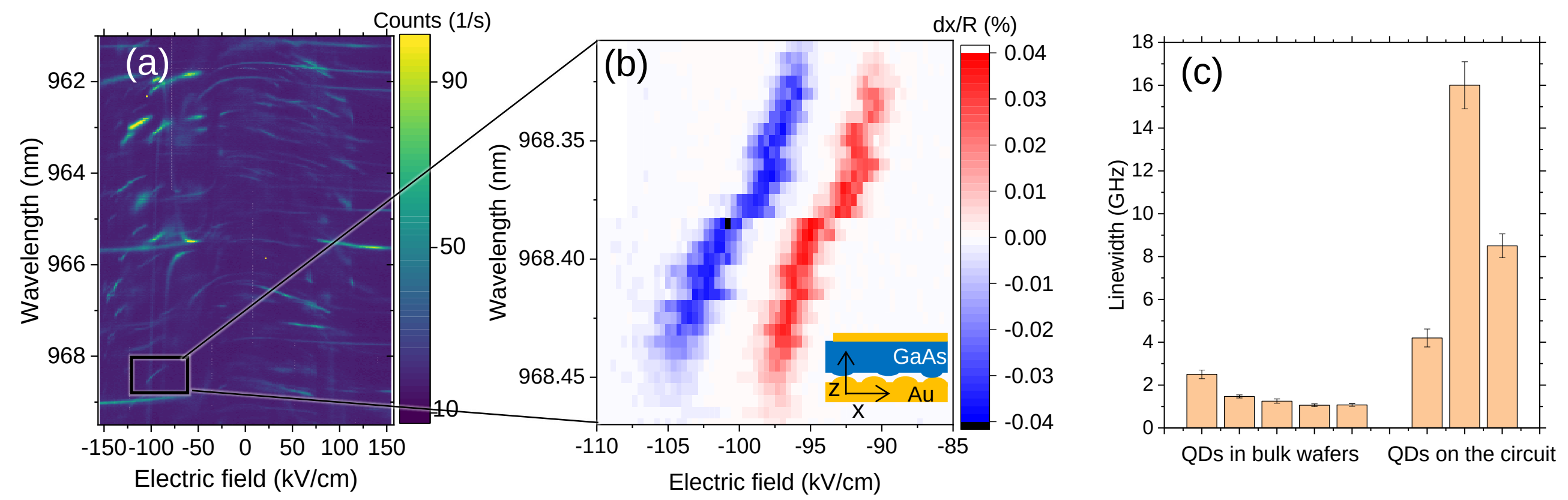

Figure 7. (a) PL from numerous QDs as a function of the bias electric field. The QDs were excited by a 780 nm laser with lower power than the absorption saturation of the QDs. The gate dependence shows clear Stark shifts and emissions from indirect transitions. (b) DR signal from a QD shown in (a). We used the 780 nm laser with the same power as for the PL measurement together with a weak resonant laser. The inset shows a schematic (not to scale) of a relatively rough back contact between the GaAs surface and the Au mirror. (c) Comparison of the QD linewidths measured by the DR spectroscopy. The error bars represent standard errors of the dispersive line fit to the spectra obtained by wavelength scans. The two linewidths and standard error bars at the most right-hand side were exceptionally estimated from the linewidths on voltage scans and their Stark shifts.

Figure 6(c) exhibits the temperature dependence of the loaded quality factors for the in-phase and out-of-phase modes at 5.25 and 7.26 GHz. With decreasing the temperature, the loaded quality factors increase to $Q_L$ = 2800 for the in-phase mode and $Q_L$ = 5300 for the out-of-phase mode. Although the frequency of the in-phase mode is shifted from the design, the quality factor still shows a reasonably high value. The corresponding decay rate is $\kappa/2\pi$ = 1.8 MHz, which is much smaller than the decay rate of an exciton in QDs, $\Gamma_{QD}/2\pi \sim 300$ MHz [38] and the expected coupling strength between an exciton and the resonator of 100s of MHz (see Section 6). The loss rate is also much lower than the typical transmon-resonator coupling strength [39]. This low-loss feature facilitates the efficient transduction between an optical and microwave qubit. If one desires to reduce the loss rate further, removing the imperfections on the nanowire resonators as well as optimizing the bias lines, as discussed in Section 2, are essential.

Besides the loss due to the bias lines, the Au layers and the QD membrane also introduce ohmic and piezoelectric losses, respectively. According to the simulations, these loss rates are on the order of ~ 1 MHz in our device. Here, the Au mirror was introduced for improving the optical coupling efficiency between an exciton and a free-space lens. As a cost of a small reduction of the outcoupling efficiency, one can eliminate the ohmic loss from the Au layers by replacing them with superconducting Al. We confirmed that the stacking process works well also on an Al layer (see Supplementary S5). The piezoelectric loss at the GaAs membrane can be minimized by reducing the size of the top gate, which would improve the internal quality factor to more than $10^4$ (see Supplementary S7) for a gate area of $\leq 1\ \mu m^2$.

We also investigated optical laser power dependence of the quality factor. We illuminated the top gate with a laser at a wavelength of 970 nm (resonant with the wavelength of QDs) and a power of 10s of nW (typical excitation power used for QD spectroscopy). Although the device substrate of silicon absorbs the laser photons efficiently, the increase in the resonator decay rate is < 1 MHz. For a better thermal noise tolerance, one should replace silicon with sapphire, which has a bandgap much larger than the resonance frequency of InAs/GaAs QDs.

## 5. Optical characterization of integrated QDs

We optically characterize QDs transferred on the coupled nanowire resonator shown in Fig. 6(b). First, we performed photoluminescence (PL) measurements while scanning the bias voltage $V_{bias}$. Each of the bias lines of the coupled nanowire circuit is connected to a voltage source and an electrical ground, respectively. The excitation laser wavelength is 780 nm, and its power (~ 0.2 nW/$\mu m^2$) is well below the absorption saturation of the QDs. The QD emission from a diffraction-limited spot through the thin top gate is collected by an objective lens with a numerical aperture (NA) of 0.7 and sent to a grating spectrometer with a charge-coupled device camera (CCD).

Figure 7(a) displays the bias electric field dependence of the PL spectrum for numerous QDs in the diffraction-limited detection spot (~ 1 $\mu m^2$). The PL map shows numerous lines because of the emissions from various QDs and charge states.

Unlike typical gate-tunable structures, the QDs are not tunnel coupled to a charge reservoir. Therefore, the charge states of the QDs are less well defined, which in turn allows for tuning the QD energy in the broader region via the DC Stark effect. On the other hand, identifying the charge states of the QD resonances by spectroscopy becomes difficult because their charge regimes overlap with each other. In Fig. 7(a), one can see clear DC Stark shifts for the overall bias range and some steep lines arising from indirect exciton recombinations, indicating the formation of the molecular states.

The overall linewidths look broader than those of QDs in an unprocessed wafer (see Supplementary S8). To investigate the linewidths more precisely, we performed resonant laser spectroscopy – differential reflection (DR) measurement [40] – by scanning the frequency of a resonant laser. We note that many PL lines visible in Fig. 7(a) are expected to originate from charged excitons because the above-band excitation accumulates electrons and holes around QDs sandwiched by the potential barriers. These lines are, in principle, not visible with the resonant spectroscopy in our structure because there are no intrinsic charge reservoirs. In order to easily find a signal in the resonant spectroscopy, we concurrently excited the sample with a 780 nm laser together with the resonant laser. This reproduces the charge configuration in the PL experiment. The excitation power of the 780 nm laser is the same as that for the PL measurement. The power of the resonant laser is ~ 100 nW/μm$^2$, which is below the saturation of the QD absorption in this device. To perform lock-in detection, we applied a bias voltage to QDs with an added kHz modulation (Amplitude: 0.1 V, Modulation frequency: 3,977 Hz). The reflection of the resonant laser and the modulated QD emission are detected by a photodiode followed by a transimpedance amplifier and a lock-in amplifier, which demodulates the signal and extracts the QD emission from the laser background.

Figure 7(b) shows the DC bias electric field dependence of the resonance wavelength for a charge state shown in Fig. 7(a). The red and blue dispersive lines separated by 0.1 V are replicas of the resonance due to the kHz modulation of the bias voltage. The DC stark shift observed here is ~ 20 μeV/(kV/cm), identifying this as a single QD resonance. The DR contrast of ~ 0.04 % (defined by the ratio of the QD emission intensity to the total reflection intensity) is orders of magnitude smaller than the typical values for gate-tunable QDs reported previously [16]. This is because the optical cavity is not optimized; namely, optical input and output coupling efficiencies are small. The resonance signal shows clear spectral fluctuations leading to a linewidth as broad as ~ 10 GHz. Here We remark that, for an unprocessed wafer, shining the above-band excitation laser does not broaden the linewidth of QDs in the resonant spectroscopy but shifts the bias voltage where the signal appears [41].

We first determine whether the broadening is due to the fabrication processes or the nature of the as-grown QD samples themselves. We investigated the linewidths for QDs on the resonator and QDs in bulk wafers by the DR measurement (Fig. 7(c)). The constitution of the QD heterostructures in the bulk wafers are the same as the one for the transferred QDs. The QD linewidths on the resonator are significantly broader than those in the bulk wafers by a factor of more than five, which indicates that the fluctuation and the broad QD linewidths on the coupled nanowire resonator are derived from the device fabrication processes. We believe that the most probable origin of the broadening is charge fluctuation arising from the relatively rough interface between the Au mirror and the QD membrane. Since the Au surface shown in Fig. 5 Step 3 has an RMS roughness of ~ 1 nm, the contacted Au-GaAs interface could produce small gaps of the same order. Electrically unshielded (100) surfaces of GaAs are a source of charge noise because of the high surface state density of > 10$^{11}$/cm$^2$ in the bandgap, determined by Terman's analysis [42].

Using a simple capacitor model, we assess whether the charge fluctuation on the poorly shielded surface can be the source of the broadening. We modelled a 130 nm-thick GaAs membrane where metal gates cover the top and back surfaces, as shown in the inset of Fig. 7(b). A laser shining on the membrane activates the midgap surface states, leading to the trapping and detrapping processes. We assume that the top surface is well contacted with the metal gate formed by an evaporation process, ensuring that the top gate screens almost all the charge fluctuations occurring on the top surface. On the other hand, we assume that the back gate and GaAs surface has a vacuum gap of $h_{\text{gap}}$ ~ 1 nm because of the roughness on the Au and GaAs surfaces. The gap induces a finite capacitance between the back gate and GaAs surface, leading to a voltage fluctuation upon charge trapping and detrapping:

$$\delta V = \frac{h_{\text{gap}}}{\varepsilon_0 S}\delta Q. \quad (2)$$

Here, $\delta V$ and $\delta Q = e\delta nS$ are voltage and charge fluctuations, $S$ is the surface area, $\varepsilon_0$ is the vacuum permittivity, $e$ is the elementary charge, and $\delta n$ represents the surface density of the fluctuating charges. By substituting the observed voltage fluctuation of $\delta V$ ~ 30 mV in Eq. (2), we estimated the charge fluctuation happening at the back contact is $\delta n$ ~ 10$^{11}$/cm$^2$. The estimated charge fluctuation does not contradict the surface state density on (100) surface, implying that the charge fluctuation at the interface of the back GaAs surface and the Au mirror can be the source of the line broadening.

One could solve the issue by fabricating a smoother back contact. For instance, one can directly deposit a Ti/Au or Al layer on the whole back surface of the QD membrane on a PDMS stamp by evaporation. Successively the membrane can be transferred on a coupled nanowire resonator followed by the top gate and sidewall lead fabrication, as shown in Fig.5

Step 4. In this case, one has to be careful not to introduce a short circuit between the metal layers deposited on the back surface and the sidewall. Instead of depositing the metal layer on the sidewall, one could fabricate an air bridge to feed microwave photons from the resonator to the top gate.

## 6. Microwave – QD coupling strength

Currently, the broad QD linewidth larger than the resonator frequency hinders us from resolving sidebands to determine the coupling strength between a microwave photon and an excitonic dipole moment experimentally. Although the charge fluctuation broadens the linewidths, the coupling strength should not be affected by this. If the charge fluctuation is suppressed, as discussed in the previous section, one should experimentally observe a significantly large single-photon coupling strength. Here we calculate how large vacuum electric fields and coupling strengths can be obtained with our device geometry. We consider the in-phase mode, which produces a strong electric field between the top and bottom gates, and neglected the bias lines by assuming the negligible leakage. For simplicity, we assume that the capacitances $C_k$ and $C_g$ are small compared with the single nanowire resonator capacitance $C_{res}$ ~ 3 fF so that the frequency shift $\delta\omega$ from the bare single resonator frequency $\omega_m$ is small; $\delta\omega/\omega_m << 1$. For the QD-coupled nanowire device, one can keep $C_g$ small by designing a small top gate area of < 1 μm$^2$. Under those assumptions, one can regard the circuit shown in Fig. 2(a) as two lambda-half resonators modelled by LC resonators with a small mutual coupling. The vacuum voltage fluctuation is then approximately given by $V_{vac} \sim \sqrt{\hbar\omega_m/2C_{res}}$. The vacuum electric field across the top and back gates are $E_{QD} = \sqrt{\varepsilon_{eff}/\epsilon_{GaAs}} V_{vac}/h_{GaAs}$, where $\varepsilon_{eff}$ is the effective dielectric constant of the nanowire resonator. The prefactor in the square root is attributed to the screening due to the high dielectric constant of GaAs. The coupling strength between the microwave field $\boldsymbol{E}_{QD}$ and an excitonic dipole moment $\boldsymbol{d}$ in a QDM reads

$$g_0 = \boldsymbol{E}_{QD} \cdot \boldsymbol{d} = \frac{|\boldsymbol{d}|}{h_{GaAs}} \sqrt{\frac{\varepsilon_{eff}}{\epsilon_{GaAs}} \frac{\hbar\omega_m^2 Z_c}{\pi}}. \quad (3)$$

Eq. (3) predicts $g_0 \propto \sqrt{Z_c}$. In particular, the single-photon coupling strength reaches $g_0/2\pi$ ~ 200 MHz when $Z_c$ ~ 5500 Ω, $\omega_m/2\pi$ ~ 10 GHz, $h_{GaAs}$ ~ 100 nm, and $|\boldsymbol{d}|/e$ ~ 4 nm. The single-photon cooperativity, in this case, is $C_0 = 4g_0/\Gamma_{QD}(\Gamma_{QD}+\kappa)$ ~ 1, leading to the internal transduction efficiency to be unity [16]. Here we assumed that the exciton recombination rate is $\Gamma_{QD}/2\pi$ ~ 300 MHz and the superconducting resonator decay rate is negligible, $\kappa << \Gamma_{QD}$, as shown in Fig. 6(c).

Although observing sidebands is currently not practical, one could perform a microwave-optical transduction experiment even with the inhomogeneously broadened QD resonance. The enhanced coupling strength of ~ 200 MHz thanks to the high-impedance resonator should improve the internal transduction efficiency by more than two orders of magnitudes compared with the design based on a coplanar waveguide resonator [16]. Since the efficiency improvement is more significant than the efficiency reduction due to the charge fluctuation, one could detect converted photons at a reasonable count rate. To this end, optimizing the optical cavity design is essential (see Supplementary S9) to improve optical input and output coupling efficiencies.

## 7. Summary and outlook

We designed and fabricated a low-loss high-impedance superconducting resonator integrated with an optically active self-assembled QDM. Thanks to the strong vacuum electric field fluctuation in the high-impedance resonator made of NbTiN nanowires, We expect a large microwave-exciton coupling strength of 100s of MHz. We demonstrated that the device features a low microwave loss rate of 1.8 MHz as well as the gate-tunability of the QD exciton resonance frequency. In order to achieve this low loss rate, we developed the transfer printing technique for self-assembled QDs, which allows for eliminating the lossy n-doped GaAs layer and piezoelectric loss from a GaAs substrate. The microwave loss rate demonstrated here is much smaller than the previously demonstrated coupling strengths between a QDM and a superconducting resonator [16], as well as the transmon-resonator coupling strength [39].

Towards efficient quantum transduction, one must improve the optical coupling efficiency limited by the two factors: (i) optical frequency mismatch between input photons and excitons because of the inhomogeneous broadening; (ii) inefficient spatial mode coupling of the QD emission to a free-space lens. To reduce the frequency mismatch, one must decrease the broad QD linewidths arising from the environmental charge noise by realizing a smoother back contact between a QD membrane and the Au mirror. With a narrower QD linewidth of << $\omega_m$, one can improve the optical outcoupling efficiency by an order of magnitude. For improved spatial mode matching, optimizing the optical cavity design is essential. By decreasing the thickness of the QD membrane to ~ 80 nm, one can resonantly couple the QD emission to the cavity mode, enhancing the coupling efficiency to a free-space lens (See Supplementary S9). These optimizations should improve the outcoupling efficiency to 10s of %. Instead of the metallic mirror, one could also employ a photonic crystal structure with a QDM, which could also be integrated into the circuit using a transfer printing technique [43, 44].

Another important requirement is to develop the deterministic positioning technique for a QDM on the circuit. Knowing the exact position of a QDM, one can reduce the size of the top gate, which enables to retain the large vacuum fluctuation and minimize the piezoelectric loss from the QD membrane, as discussed in Section 4. Furthermore, the positioning technique allows us to choose a target QDM prior to the integration [37]. This is beneficial to couple a QDM to a superconducting circuit with a matched microwave frequency, also essential for efficient transduction [16].

The obtained insights and developed techniques in this paper should be beneficial not only for quantum transducers but also for other advanced devices. For instance, the transfer printing technique for self-assembled QDs with gate tunability could be applicable for other advanced photonic devices based on QDs. On the other hand, our coupled high-impedance resonators with DC bias lines can possess other materials instead of QDs at the position of the coupling electrodes. One could think of coupling the large electric vacuum fluctuation with other materials' degrees of freedom to modify material properties.

## Acknowledgements

This publication was produced within the scope of the National Research Programme NCCR QSIT, which was funded by the Swiss National Science Foundation. We would like to thank the Swiss National Science Foundation for their financial support. We acknowledge helpful discussion on superconducting circuits with Benedikt Kratochwil under the supervision of Klaus Ensslin. We thank Mihai Gabureac for the help and discussions about superconducting circuit fabrication under the supervision of Andreas Wallraff. We thank Clemens Todt for superconducting film characterization and Stefan Fält for QD growth under the supervision of Werner Wegscheider. We acknowledge valuable discussion about fabrication and simulations with Emre Togan. We appreciate the indispensable supports for the project from Ataç İmamoğlu. We thank the Cleanroom Operations Team of the Binnig and Rohrer Nanotechnology Center (BRNC) and FIRST laboratory for their help and support. The authors declare no competing interests.

## Author contributions

Y.T. performed sample fabrication, measurement, and analysis. Y.T. and M.K. supervised the project and wrote the manuscript.

## Data availability statement

The data is available at DOI xxx.

# Supplementary: Low-loss high-impedance circuit for quantum transduction between optical and microwave photons

**Yuta Tsuchimoto and Martin Kroner**

Department of Physics, ETH Zurich, Zurich, Switzerland

E-mail: tyuta@phys.ethz.ch

## S1. Fabrication of a coupled nanowire resonator

We fabricated a coupled nanowire resonator without QDs on a high-resistivity silicon substrate (~ 20 kΩcm). First, we sputtered a thin NbTiN film (15 nm) on a silicon substrate. The sheet resistance of the film is 378 Ω/sq at room temperature. We patterned nanowires with a width of ~ 70 nm using electron beam lithography (EBL) followed by reactive ion etching (RIE) in $SF_6/O_2$ plasma. The design parameters of the nanowire resonators and bias lines are mentioned in Section 2.1 in the main text. Feedlines were pattered by photolithography and RIE.

Figure S1(a) shows optical and SEM images of a fabricated coupled nanowire resonator and bias lines. We designed the electrode gap shown in the inset as small as 200 nm for this test device to reproduce the capacitance (~ 1 fF) of a QD-coupled nanowire device with additional metal gates. Each nanowire resonator is connected to a bias line at the field node, forming "T" junctions (see the inset). The roughness of the silicon substrate seen in the SEM images is due to residues of a positive tone electron beam resist, which do not affect the quality factor of the resonator at the level of our interest. We removed those residues by increasing the development time and obtained a flat etched surface, as shown in the main text.

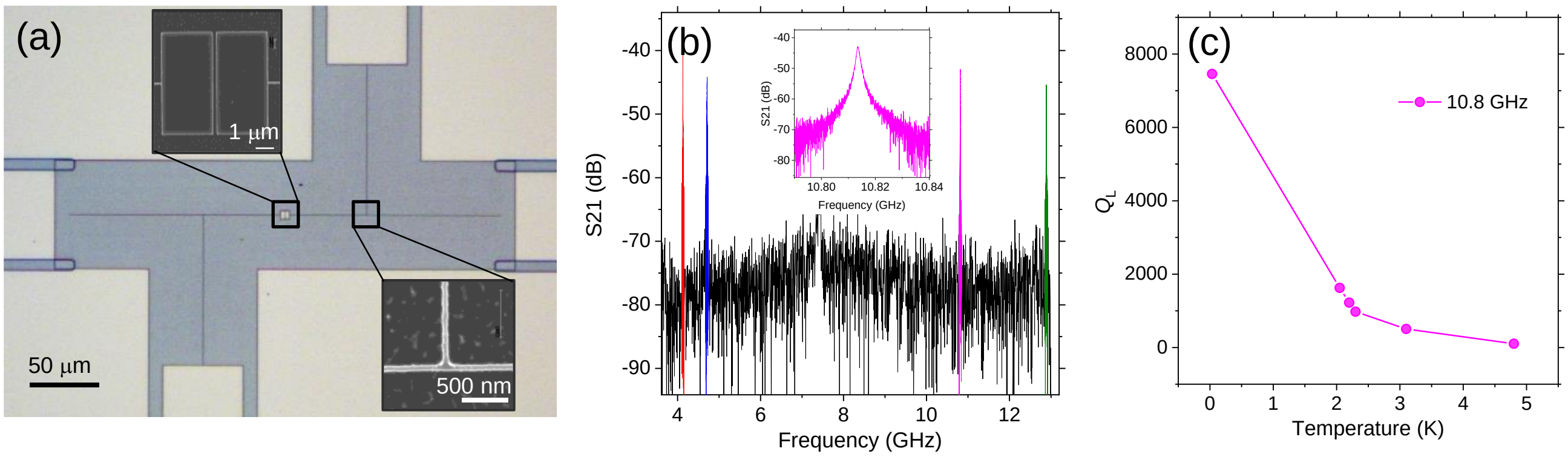

Figure S1. (a) Images of a fabricated coupled nanowire resonator. The insets are SEM images of the coupling electrodes and the T junction. (b) measured |*S*21| at 55 mK. The four resonant modes observed here correspond to the simulated modes in Fig. 2(c) in the main text. Each resonant mode is highlighted with different colours. The inset shows the enlarged spectrum range of the in-phase mode (magenta). (c) Temperature dependence of the loaded quality factor of the in-phase mode.

We performed transmission measurements for the sample at various temperatures. Since we designed $C_k$ to be relatively large, the transmission is detectable even at ~ 4 K, where the resonator quality factor is typically low. The input microwave drive from a network analyzer was attenuated such that it does not excite the nonlinearity of the resonator (see Supplementary S3). The transmitted signal was amplified by ~ 60 dB at room temperature and fed back into the network analyzer. Figure S1(b) shows measured $|S_{21}|$ at 35 mK. As expected from Fig. 2(c), there are four peaks, and those resonance frequencies agree well with the simulated resonance frequencies at the kinetic inductance of 83 pH/sq. The inset of Fig. S1(b) is the enlarged spectral range for the in-phase mode, and the temperature dependence of the loaded quality factor for the corresponding mode is plotted in Fig. S1(c). The quality factor at 10.81 GHz reaches $Q_L$ ~ 8000, limited by the external decay rate because of the relatively large $C_k$. Unexpectedly, the other modes, which should not be protected from the leakage, show also high-quality factors. The reason for this unexpectedly high-quality factor is currently under investigation.

## S2. QD wafer structure

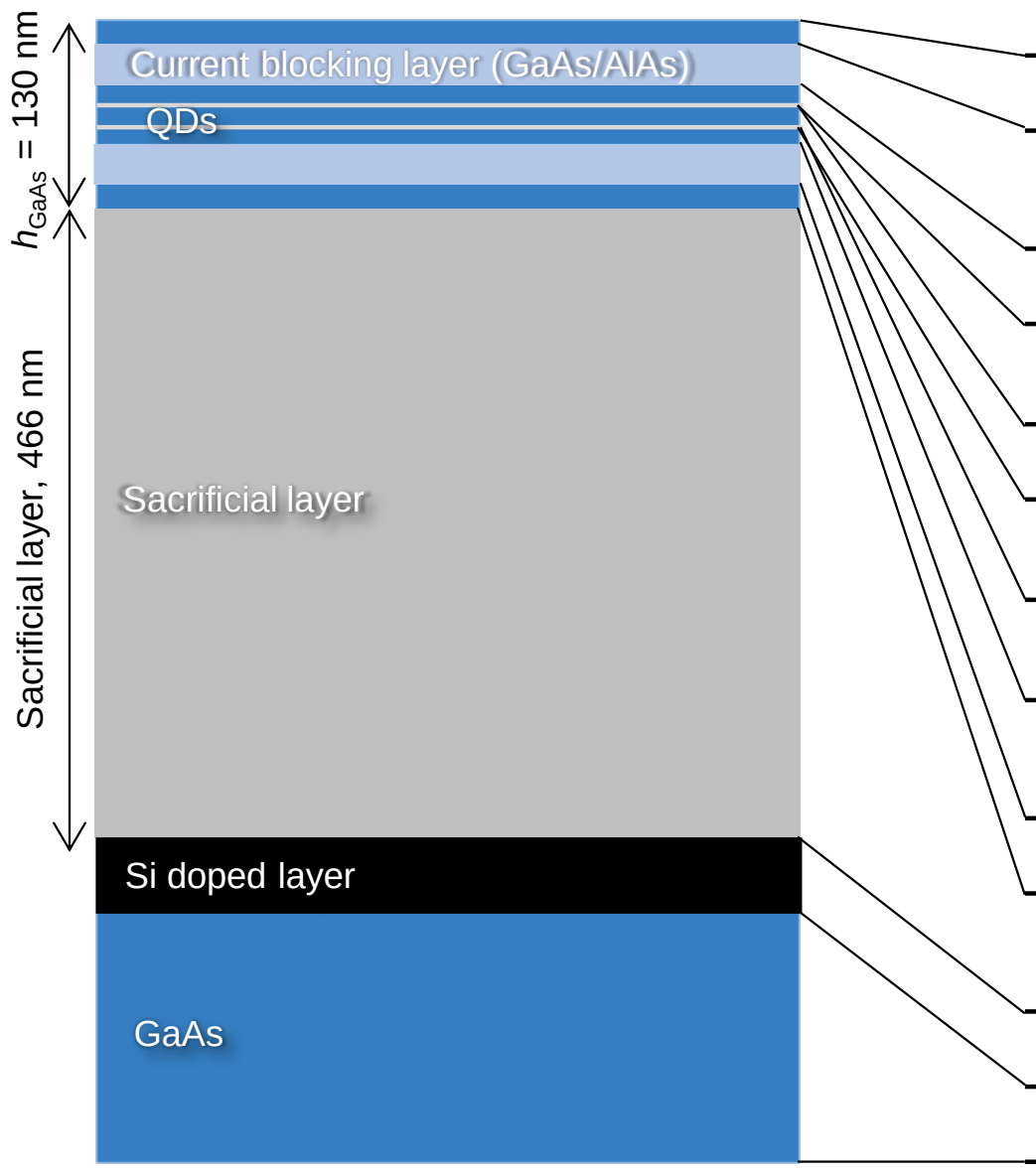

| | Material | Thickness (nm) | |
|---|---|---|---|
| Protection | GaAs | 17 | |
| Current block | GaAs<br>AlAs | 1.9<br>1.9 | × 8 |
| Cap | GaAs | 14 | |
| Top dots | InAs/GaAs | - | |
| Spacer | GaAs | 12 | |
| Bottom dots | InAs/GaAs | - | |
| Cap | GaAs | 10 | |
| Current block | GaAs<br>AlAs | 1.9<br>1.9 | × 8 |
| Protection | GaAs | 17 | |
| Sacrificial layer | AlAs<br>GaAs | 2.1<br>0.53 | × 177 |
| Doped layer | Si:GaAs | 56 | |
| Buffer | GaAs | 188 | |

Figure S2. Structure and thicknesses of MBE grown layers. QDMs are embedded between current blocking layers made of AlAs/GaAs superlattices. A sacrificial layer is made of Al-rich AlAs/GaAs superlattice under the QDMs and the current blocking layers. A doped layer underneath the current blocking layer allows for characterizing bias voltage dependence of the QD resonances in a bulk wafer.

## S3. Input power dependence of a coupled nanowire resonator

We investigated the input power dependence of the resonance of a coupled nanowire resonator. The sample used in this experiment is identical to the one shown in Fig. S1. We input a signal from a network analyzer with various attenuations and fed the transmission back to the analyzer after amplified by ~ 60 dB at room temperature. We calculated input powers by considering the attenuators on the input path and the cable losses outside of a cryostat. Figure S3 displays $|S_{21}|$ for various input powers at 55 mK, showing a nonlinear response of the resonance when the input power is -77 dB. In the main text, we performed microwave experiments with a smaller input power so that the input drive does not excite the nonlinearity.

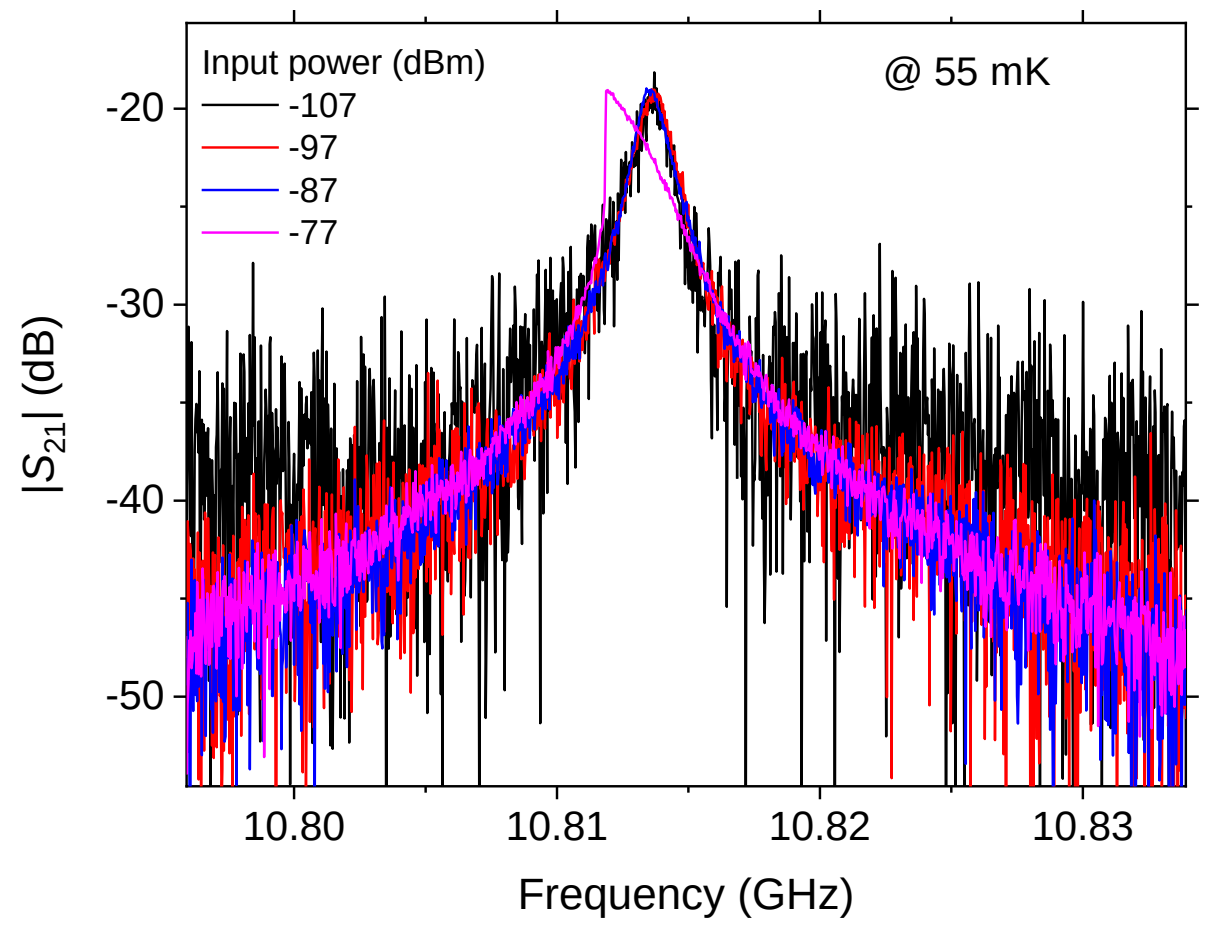

Figure S3. Input power dependence of $|S_{21}|$ for the coupled nanowire resonator shown in Fig. S1. We changed the input power from -107 to -77 dBm and measured the transmission at 55 mK.

## S4. Back surface cleaning of a QD membrane

After picking a QD membrane up with an elastomer stamp made of polydimethylsiloxane (PDMS), we first cleaned the back surface of the QD membrane by 3 % HF for 1 min. Figure S4(a) displays atomic force microscope (AFM) topography images of the back surface of the QD membrane after the HF cleaning, showing many island structures with a height of ~ 3 nm. Those structures could be residues such as Aluminum fluoride produced during the HF etching process in Fig. 4 Step 4. The root-mean-square (RMS) roughness of the surface is ~ 1.5 nm, corresponding to the previously reported value [1]. To remove those residues, we cleaned the surface by digital etching consisting of surface oxidation by 30% $H_2O_2$ followed by acid cleaning with 16 % HCl for 1 min. In addition, we immersed the membrane into 3 % HF solvent again. The surface after the cleaning process looks much cleaner, as shown in Fig. S4(b). There are only a few islands on the surface, and the RMS roughness is ~ 0.5 nm. The mechanism of removing the contaminations is currently under investigation. An acidic solvent may etch out oxidized residues by $H_2O_2$, or the reaction of the residues with HCl itself could play a role.

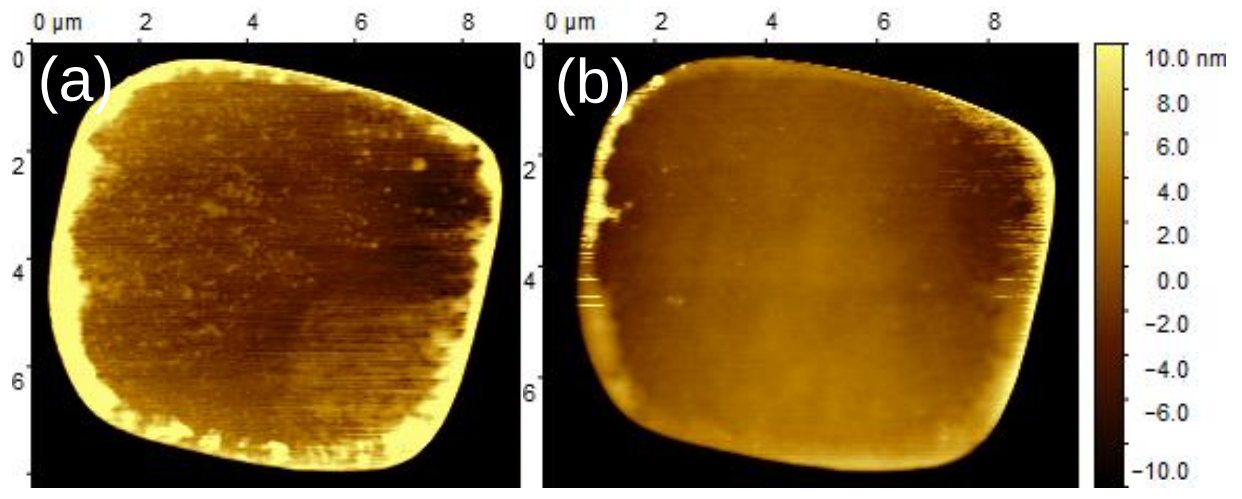

Figure S4. Topography images of the back surface of a QD membrane picked up by a PDMS stamp. The images were taken by an AFM in tapping mode. (a) After the cleaning with 3 % HF for 1 min. (b) After additional cleaning by digital etching with 30% $H_2O_2$ and 16 % HCl followed by cleaning with 3 % HF again.

## S5. Various stacking conditions

In order to investigate the importance of some parameters for the stacking process, we tried to transfer QD membranes under different conditions: (i) several treatment conditions just before the stacking shown in Fig. 5 (without the treatment, with HF treatment, or with HCl treatment); (ii) different accepter materials on which we stack a QD membrane (Au, Al, or NbTiN); (iii) two-gap lengths between the electrodes (1.0 or 1.5 μm), which changes the size of the contact area. For each condition, we tried stacking with multiple clean membranes at least several times for every membrane. The results are summarized in Table S1. The condition labelled as "Succeed" never failed within the number of trials, whereas those labelled as "failure" never succeeded.

When the treatment was 3 % HF and the gap length was 1.0 μm, the stacking process succeeded regardless of the accepter materials. We note that the RMS surface roughness of the Au layer is ~ 1 nm, and if the roughness is larger than that, the stacking process may fail.

Once we increased the gap length from 1.0 to 1.5 μm while keeping the other parameters, the stacking process did not succeed. This is because the adhesive force between the back GaAs surface and the acceptor material became small due to the small contact area.

We realized that the HF treatment just before the stacking process shown in Fig. 5 Step 2 is essential for the stacking process. We prepared QD membranes treated with 3% HF and stored them in an Ar atmosphere for about a week. After that, we tried to stack the membranes on a circuit without additional treatment. The stacking process did not succeed with these membranes; however, it worked deterministically as soon as we treated the membranes with 3% HF for 1 min just before the stacking process.

We also tried stacking some membranes just after the treatment with 16% Hydrochloric acid (HCl), and it turned out that the stacking process did not work, contrary to the case of the HF treatment. Although the origin of the adhesiveness between the GaAs back surface and the accepter materials is not clear, the higher success rate with the HF treatment could be related to the surface passivation of GaAs. Further investigation is beyond the scope of this paper.

| Conditions | Accepter surface | Success/Failure |
|---|---|---|
| HF treatment, Gap 1.0 μm | Au / Al / NbTiN | Success |
| HF treatment, Gap 1.5 μm | Au / Al | Failure |
| No treatment, Gap 1.0 μm | Au | Failure |
| HCl treatment, Gap 1.0 μm | NbTiN | Failure |

Table S1. Summary of the various stacking conditions. We performed the stacking process under different surface treatments, two different gap lengths, and several contact materials.

## S6. Microwave resonances of QD-coupled nanowire devices

We measured $|S_{21}|$ of a QD-coupled nanowire resonator with a similar structure in a Helium bath cryostat at ~ 2 – 4 K. The sample has basically the same design (see the main text for the detailed design) but has a slightly wider nanowire width (~ 50 nm) than that shown in Fig. 6.

We identified the resonances derived from the coupled nanowire resonator by observing the temperature-dependent resonance frequency shifts due to the change of kinetic inductance. The top panel in Fig. S5(a) shows the $|S_{21}|$ of the resonator measured at ~ 2 – 4 K. There are four peaks at 2.8, 3.9, 7.9, and 10.7 GHz corresponding to the simulated spectrum in the bottom panel. Here we performed the simulation by using ANSYS HFSS. The resonances at 7.9 and 10.7 GHz are in-phase and out-of-phase modes, respectively, and the resonances at the shorter frequencies are associated with the bias lines. The transmissions of the resonances at 2.8 and 10.7 GHz are significantly weaker because of the low internal quality factors due to the leakage expected from the simulation. However, the resonance at 3.9 GHz shows an unexpectedly high transmission and loaded quality factor, even though the simulation indicates significant damping. Currently, the reason is not apparent. The in-phase mode at 7.9 GHz is protected from leakage, showing the high transmission and loaded quality factor. Figure S5(b) shows the temperature dependence of the loaded quality factor for the in-phase mode. The loaded quality factor increases to ~ 900, corresponding to a decay rate of 9 MHz at 2.2 K, which is almost the same as the one shown in the main text. The transmission of the in-phase mode at 2.2 K is also a similar amount. The loaded quality factor shown here is currently limited by the temperature and should improve to the same order as shown in the main text by cooling down to < 1 K.

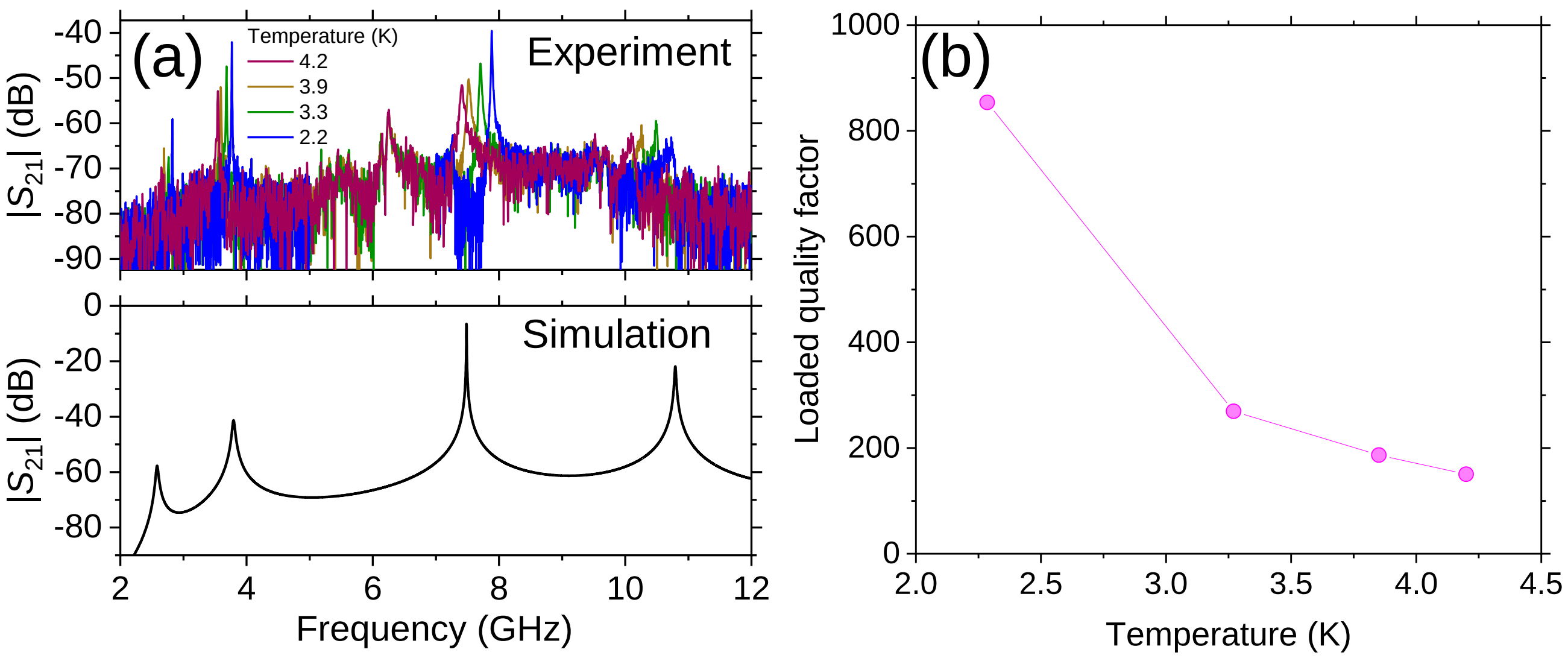

Figure S5. (a) Measured (top panel) and simulated (bottom panel) $|S_{21}|$ of a QD-coupled nanowire resonator with a nanowire width of ~ 50 nm. (b) Temperature dependence of the loaded quality factor for the in-phase mode at 7.9 GHz.

## S7. Piezoelectric loss of the QD-nanowire resonator device

We simulated piezoelectric loss at a QD membrane made of mainly GaAs using a commercial finite element solver COMSOL. We modelled the QD membrane as a cuboid membrane made of pure GaAs, of which the dielectric constant is $\varepsilon_{\text{GaAs}} = 12.8$. We set a partial part of the top surface and the whole bottom surface of the membrane as perfectly conductive layers. This structure mimics the capacitively coupled top and bottom gates of a QD-coupled nanowire device. We applied an AC voltage with a frequency of 10 GHz on the top gate and connected the bottom gate to the ground, leading to a vertical AC electric field oriented to (100) direction (see the cross-sectional schematic in Fig. S6(a)). The AC electric field excites an acoustic mode because of the piezoelectricity of GaAs. The displacement amplitude of the acoustic mode is ~ $10^{-17}$ m when the applied voltage is ~ 10 μV corresponding to the vacuum fluctuation of our high-impedance resonator. The displacement direction is perpendicular to the electric field direction.

We calculated all the dissipated energies as a function of the GaAs thickness. We estimated the piezoelectric loss ratio, $r_{\text{loss}}$, by normalizing the dissipated energies by capacitive energy stored in the top and bottom gates $W_g = 1/4C_gV^2$. Figure S6(b) exhibits the calculated $r_{\text{loss}}$. The loss ratio takes local maxima when the acoustic mode is efficiently generated by resonating with the top and bottom gates. Except for the resonant cases, $r_{\text{loss}}$ is less than 1 %.

From $r_{\text{loss}}$, we calculated resonator quality factors $Q_{\text{piezo}}$ exclusively limited by the piezoelectric loss. The quality factor of the QD-coupled nanowire resonator can be estimated by the following equation:

$$Q_{\text{piezo}} = 2\pi \frac{W_g + 2W_{\text{res}}}{r_{\text{loss}} W_g} = 2\pi \frac{C_g + 4C_{\text{res}}}{r_{\text{loss}} C_g}. \qquad \text{(S1)}$$

The terms $W_g = 1/4C_gV^2$ expresses capacitive energy stored between the top and bottom gates of the QD membrane, and $W_{\text{res}} = 1/2C_{\text{res}}V^2$ represents the total (inductive and capacitive) energy stored in a nanowire resonator, respectively [2]. Here $C_g$ is the capacitance between the top and back gates, and $C_{\text{res}}$ is the capacitance of a single nanowire resonator. Eq. (S1) indicates that the quality factor of the high-impedance resonator is more susceptible to the loss at the QD position than that of a typical coplanar resonator because of the lower $C_{res}$ of the high-impedance resonator. This loss sensitiveness of high-impedance resonators requires a design that eliminates loss channels of a coupled medium as a cost of the large coupling strength.

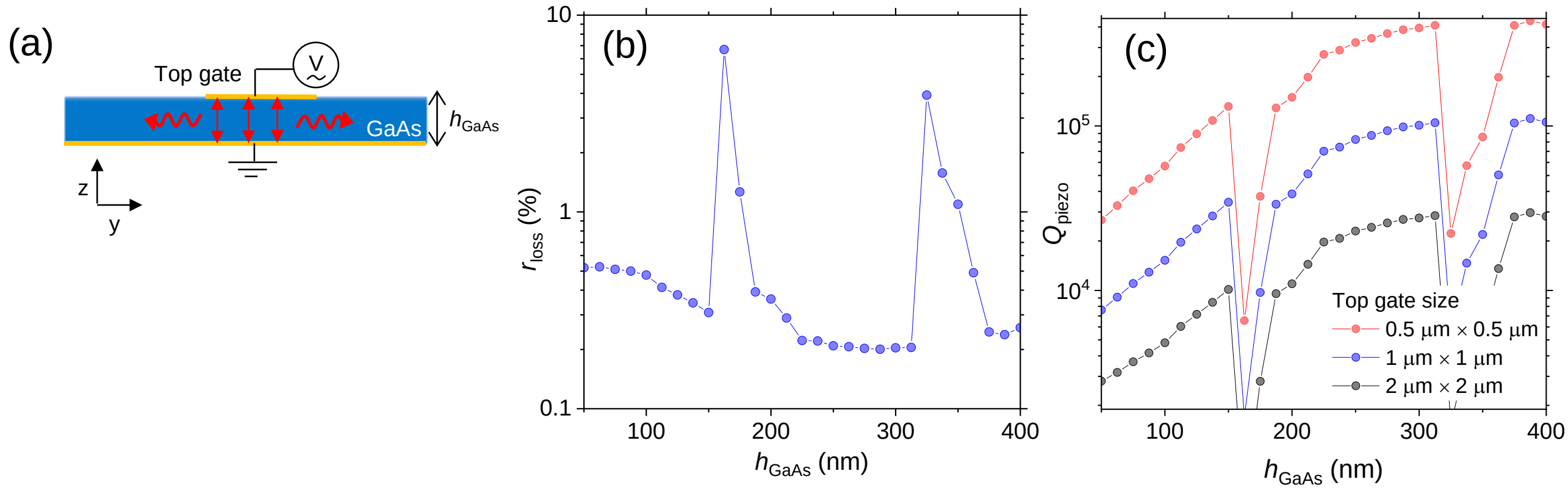

Figure S6. Simulation for piezoelectric loss at the QD membrane as a function of the membrane thickness $h_{\text{GaAs}}$. (a) Cross-sectional schematic of a GaAs membrane sandwiched by a top and back gates. The AC electric field between the gates generates an acoustic mode propagating to the edge of the membrane. (b) Simulated energy loss against the capacitive energy stored in the gate structure. (c) Simulated quality factors defined by the ratio of the stored energy in the device to the acoustic dissipation. The red, blue, and black points correspond to the cases for the top gate sizes of 0.5 μm × 0.5 μm, 1 μm × 1 μm, and 2 μm × 2 μm respectively. The gate size of 2 μm × 2 μm corresponds to the one used in the experiment.

Figure S6(c) shows the calculated $Q_{piezo}$ as a function of the thickness with the gate sizes from 0.5 μm × 0.5 μm to 2 μm × 2 μm. The quality factors take local minima when $r_{loss}$ takes local maxima in Fig. S6(b). When $h_{GaAs}$ becomes smaller, $Q_{piezo}$ decreases because of the larger $C_g$ in Eq. (S1). This is because more energy is stored between the top and back gates rather than the nanowire resonators, leading to a larger piezoelectric loss at the QD position. Contrary, $Q_{piezo}$ increases by reducing the top gate size because of the smaller $C_g$. At $h_{GaAs}$ ~ 130 nm, corresponding to our design, the quality factor exclusively limited by the piezoelectric effect is $Q_{piezo}$ ~ 7000 for the gate sizes of 2 μm × 2 μm. For a smaller gate size of 0.5 μm × 0.5 μm, the quality factor improves to $Q_{piezo} \sim 9 \times 10^4$. This tells us that fabricating a smaller gate area is essential to improve the quality factor. To this end, one can fabricate the top gate only above a target QDM by first fabricating position markers for a target QDM in a bulk wafer [3], followed by the transfer printing of the QDM with the markers. The gate area can be minimized by depositing the top gate only around the target QDM with respect to the markers. One could also think of incorporating QDs into a phononic structure with a bandgap around the microwave resonance frequency in order to reduce the piezoelectric loss.

## S8. Optical properties of QDs in a bulk wafer

Figures S7 (a) and (b) show PL maps of QDs on the coupled nanowire resonator and QDs in a bulk wafer, respectively. We excited excitons in QDs by 780 nm laser photons with a power sufficiently below the saturation of the QD absorption. Both PL maps exhibit similar charge plateaus, DC Stark shifts, and emissions from indirect excitons. The PL from the QDs on the resonator show relatively broader linewidths than those of the bulk wafer. Especially, the linewidths of the indirect emissions are significantly broader than those for the bulk wafer because the indirect exciton energies are much more sensitive to charge fluctuation in the environment due to the larger excitonic dipole moment.

Because of the fluctuation, the differential reflection (DR) signal from an exciton in a QD on the resonator shows a linewidth of more than 10 GHz and significant energy fluctuation, as shown in the main text. To exclude the possibility that the broadening is the nature of the as-grown QDs themselves, we also checked DR signals from QDs in a bulk wafer. The QDs were driven by a weak resonant laser together with a weak 780 nm laser, which is the same excitation condition for the QD measurement on the resonator. Figure S7(c) shows the DR signals from several QDs in the bulk wafer, exhibiting clear DC Stark shifts without the charge fluctuation. The linewidth of the signal shown in the inset is as narrow as ~ 1 GHz. These results indicate that the origin of the fluctuation is related to the fabrication processes for the QD-coupled nanowire device. As discussed in the main text, this charge fluctuation is most arguably due to the rough interface between the GaAs membrane and the Au mirror of the device.

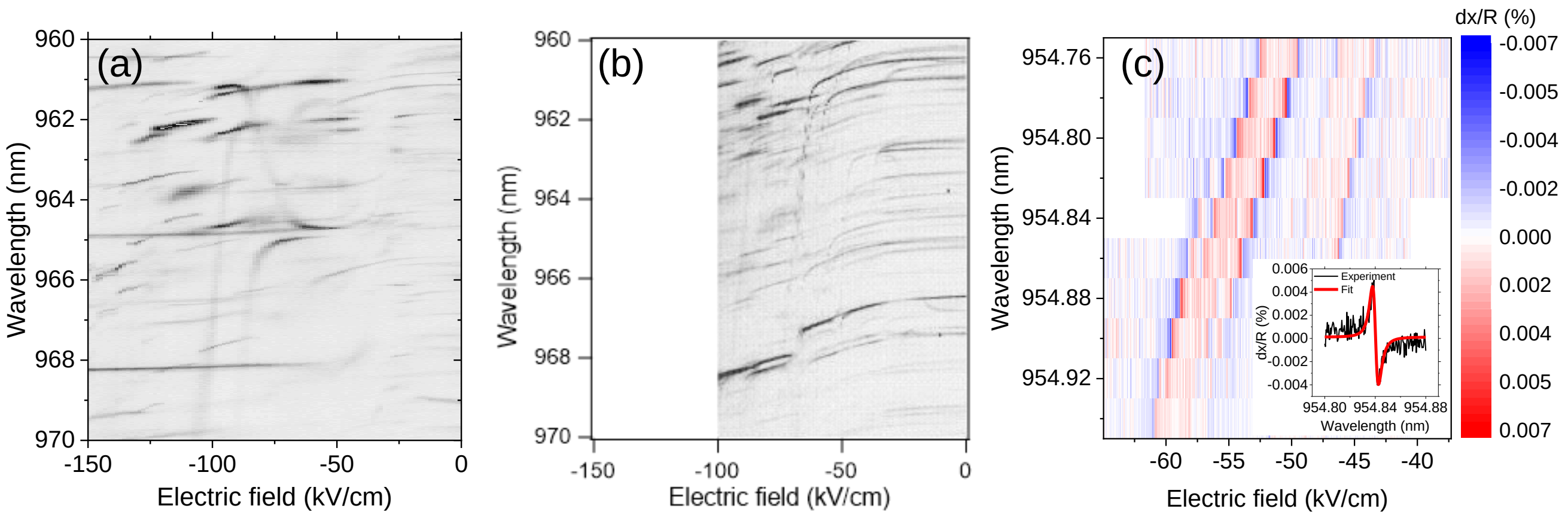

Figure S7. (a, b) PL maps for (a) QDs on the coupled nanowire resonator shown in the main text and (b) QDs in a bulk wafer. The heterostructures of both QDs have the same structures. (c) DR signals from several QDs in a bulk wafer. The signals were measured under 780 nm excitation together with a resonant laser. The inset shows a wavelength scan at a fixed bias voltage.

## S9. Optical outcoupling

In the main text, we show that the DR signal from the QD is weak (~ 0.04 %) because of the suboptimal optical cavity. To improve the optical coupling efficiency, we investigated the collection efficiency at a first lens as a function of the cavity thickness using finite-element simulation with COMSOL. The inset of Figure S8(a) displays a schematic of the simulated structure. A single emitter in the GaAs membrane is coupled to the optical cavity formed by the 15 nm-thick Au top layer and the 100 nm-thick bottom Au mirror. The single emitter was modelled by an electrical current dipole moment oriented parallel to the surface. The dipole moment emits light at 970 nm into the vacuum through the thin top Au layer, which was collected over the area corresponding to NA = 0.7. The light is also emitted into the silicon substrate through the thick Au layer and a 15 nm-thick NbTiN. The Au and NbTiN layers absorb some amount of light because of the imaginary parts of the refractive indices. For the refractive indices, we used the data of Johnson and Christy [4] for Au and Banerjee et al. [5] for NbTiN, and used a refractive index of 3.5 for GaAs and Si. We defined the collection efficiency $\eta_0$ as the ratio of the detected power in the vacuum at NA = 0.7 to the whole power emitted from the dipole.

Figure S8(a) shows $\eta_0$ as a function of the optical cavity length $h_{\mathrm{GaAs}}$ (i.e., the thickness of the QD membrane). When the optical cavity length is 130 nm, corresponding to the QD membrane thickness used in the experiment, the calculated $\eta_0$ is as low as ~ 4 %. At this cavity length, the emission does not resonate with the cavity. Hence, a considerable amount of the emission from the QD is scattered into the silicon substrate due to the substantial refractive index contrast between GaAs and vacuum, as shown in the emission pattern in Fig. S8(b). With decreasing the length to ~ 80 nm, the efficiency $\eta_0$ becomes higher and reaches ~ 14 %. In this case, the emission at 970 nm is resonant with the asymmetric optical cavity; therefore, optical photons efficiently leak out from the thin top Au layer, as shown in Fig. S8(c). We note that the actual collection efficiency at off-resonant conditions can be overestimated because the field mode at the first lens is far from the Gaussian distribution.

QD wafers grown by MBE have some thickness and QD density gradients. In order to obtain an 80 nm-thick membrane with a reasonable amount of QDMs, one needs to optimize the growth process such that the regions of the correct thickness, QD densities and exciton energies are overlapped on a wafer.

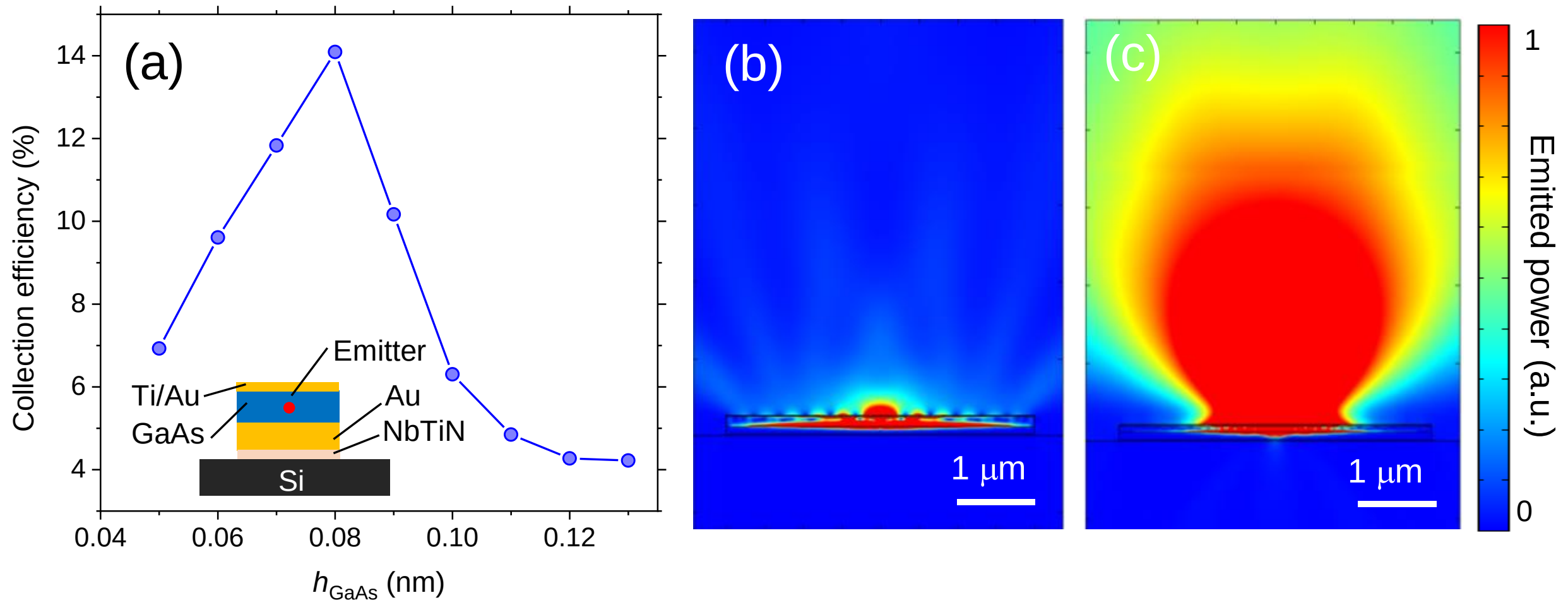

Figure S8. Simulations for the emission from a QD on the nanowire resonator. (a) Collection efficiency at a first lens with NA = 0.7. The inset shows the simulated structure. The efficiency was calculated as the ratio of the detected power at the position of the lens to the total emission power. (b, c) show distributions of the emitted power for the thicknesses of the QD membrane for (b) $h_{\mathrm{GaAs}}$ = 130 nm, same as the thickness of the QD membrane shown in the main text, and (c) $h_{\mathrm{GaAs}}$ = 80 nm.